\documentclass[twocolumn,times,twocolappendix]{aastex631}

\usepackage{color}
\definecolor{br}{rgb}{0.8,0.25,0.33}

\usepackage{amsmath}

\shorttitle{Resolved CO~$J$~=~2--1 and dust emissions in $z=1.46$ cluster galaxies}
\shortauthors{Ikeda et al.}

\graphicspath{{./}{figures/}}

\begin{document}

\title{\Large{High-resolution ALMA Study of CO~$J$~=~2--1 line and Dust Continuum Emissions in Cluster Galaxies at $z=1.46$}}

%Authors & Affiliations
\author[0000-0002-2634-9169]{Ryota Ikeda}
\correspondingauthor{Ryota Ikeda}
\email{ryota.ikeda@grad.nao.ac.jp}
\affiliation{Department of Astronomy, School of Science, SOKENDAI (The Graduate University for Advanced Studies), 2-21-1 Osawa, Mitaka, Tokyo 181-8588, Japan}
\affiliation{National Astronomical Observatory of Japan, 2-21-1 Osawa, Mitaka, Tokyo 181-8588, Japan}

\author[0000-0001-9728-8909]{Ken-ichi Tadaki}
\affiliation{Department of Astronomy, School of Science, SOKENDAI (The Graduate University for Advanced Studies), 2-21-1 Osawa, Mitaka, Tokyo 181-8588, Japan}
\affiliation{National Astronomical Observatory of Japan, 2-21-1 Osawa, Mitaka, Tokyo 181-8588, Japan}

\author[0000-0002-2364-0823]{Daisuke Iono}
\affiliation{Department of Astronomy, School of Science, SOKENDAI (The Graduate University for Advanced Studies), 2-21-1 Osawa, Mitaka, Tokyo 181-8588, Japan}
\affiliation{National Astronomical Observatory of Japan, 2-21-1 Osawa, Mitaka, Tokyo 181-8588, Japan}

\author[0000-0002-2993-1576]{Tadayuki Kodama}
\affiliation{Graduate School of Science, Tohoku University, 6-3 Aramaki Aza-Aoba, Sendai, Miyagi 980-8578, Japan}

\author[0000-0001-6251-3125]{Jeffrey C. C. Chan}
\affiliation{Private Address, London, SE8, United Kingdom}

\author[0000-0001-6469-8725]{Bunyo Hatsukade}
\affiliation{Institute of Astronomy, Graduate School of Science, The University of Tokyo, 2-21-1 Osawa, Mitaka, Tokyo 181-0015, Japan}

\author[0000-0002-9321-7406]{Masao Hayashi}
\affiliation{National Astronomical Observatory of Japan, 2-21-1 Osawa, Mitaka, Tokyo 181-8588, Japan}

\author[0000-0001-9452-0813]{Takuma Izumi}
\affiliation{Department of Astronomy, School of Science, SOKENDAI (The Graduate University for Advanced Studies), 2-21-1 Osawa, Mitaka, Tokyo 181-8588, Japan}
\affiliation{National Astronomical Observatory of Japan, 2-21-1 Osawa, Mitaka, Tokyo 181-8588, Japan}

\author[0000-0002-4052-2394]{Kotaro Kohno}
\affiliation{Institute of Astronomy, Graduate School of Science, The University of Tokyo, 2-21-1 Osawa, Mitaka, Tokyo 181-0015, Japan}
\affiliation{Research Center for the Early Universe, The University of Tokyo, 7-3-1 Hongo, Bunkyo, Tokyo 113-0033, Japan}

\author[0000-0002-0479-3699]{Yusei Koyama}
\affiliation{Department of Astronomy, School of Science, SOKENDAI (The Graduate University for Advanced Studies), 2-21-1 Osawa, Mitaka, Tokyo 181-8588, Japan}
\affiliation{Subaru Telescope, National Astronomical Observatory of Japan, 650 North A’ohoku Place, Hilo, HI 96720, U.S.A.}

\author[0000-0003-4442-2750]{Rhythm Shimakawa}
\affiliation{National Astronomical Observatory of Japan, 2-21-1 Osawa, Mitaka, Tokyo 181-8588, Japan}

\author[0000-0002-3560-1346]{Tomoko L. Suzuki}
\affiliation{Kavli Institute for the Physics and Mathematics of the Universe (WPI), The University of Tokyo Institutes for Advanced Study, The University of Tokyo, Kashiwa, Chiba 277-8583, Japan}

\author[0000-0003-4807-8117]{Yoichi Tamura}
\affiliation{Division of Particle and Astrophysical Science, Graduate School of Science, Nagoya University, Furo-cho, Chikusa-ku, Nagoya, Aichi 464-8602, Japan}

\author[0000-0002-4937-4738]{Ichi Tanaka}
\affiliation{Subaru Telescope, National Astronomical Observatory of Japan, 650 North A'ohoku Place, Hilo, Hawaii, 96720, USA}

\begin{abstract}
We present new Atacama Large Millimeter/submillimeter Array (ALMA) results obtained from spatially resolved CO~$J$~=~2--1 line ($0\farcs4$ resolution) and 870 $\mu$m continuum ($0\farcs2$ resolution) observations of cluster galaxies in XMMXCS J2215.9-1738 at $z=1.46$. Our sample comprises 17 galaxies within $\sim0.5$ Mpc ($0.6R_{200}$) of the cluster center, all of which have previously been detected in the CO~$J$~=~2--1 line at a lower resolution. The effective radii of both the CO~$J$~=~2--1 line and 870 $\mu$m dust continuum emissions are robustly measured for nine galaxies by modeling the visibilities. We find that the CO~$J$~=~2--1 line emission in all of the nine galaxies is more extended than the dust continuum emission by a factor of $2.8\pm1.4$. 
We investigate the spatially resolved Kennicutt-Schmidt (KS) relation in two regions within the interstellar medium of the galaxies. The relation for our sample reveals that the central region ($0<r<R_{e,{\rm 870\mu m}}$) of galaxies tends to have a shorter gas depletion timescale, i.e., a higher star formation efficiency, compared to the extended region ($R_{e,{\rm 870\mu m}}<r<R_{e,{\rm CO}}$). Overall, our result suggests that star formation activities are concentrated inside the extended gas reservoir, possibly resulting in the formation of a bulge structure. We find consistency between the ALMA 870 $\mu$m radii of star-forming members and the Hubble Space Telescope/1.6 $\mu$m radii of passive members in a mass-size distribution, which suggests a transition from star-forming to passive members within $\sim0.5$ Gyr. In addition, no clear differences in the KS relation nor in the sizes are found between galaxies with and without a close companion. 
\end{abstract}

%Unified Astronomy Thesaurus concepts?

\received{February 27, 2022}
\accepted{May 3, 2022}

\section{Introduction}\label{sec:Introduction}

%P1:銀河進化と銀河団、一般論
In the local universe, galaxies show a variety of morphologies and star formation properties, provoking questions about their evolution over the past 13.8 Gyr. One of the key factors controlling these properties is the environment in which the galaxies reside. A large fraction of the galaxies in nearby clusters are ellipticals (e.g. \citealp{1980ApJ...236..351D}), red (e.g. \citealp{2009MNRAS.393.1324B}), and passive (quiescent) galaxies (e.g. \citealp{2004MNRAS.353..713K}; \citealp{2005ApJ...629..143B}), which is in stark contrast to low-density (field) environments. Although such environmental dependencies of galaxies in the local universe have been known for decades, the whole picture of the evolution of massive ellipticals is not yet conclusive. Therefore, observations of high-redshift clusters, as the progenitors of nearby clusters, are important objects for further understanding how massive galaxies form and evolve down to the present-day universe.

%P2:銀河団中の星形成のphaseの話
With the aid of expansive multiwavelength observations and simulations, we now have a general understanding of cluster evolution, to some extent. Broadly, the evolutionary phase of the star formation activity of galaxy clusters can be divided into three epochs \citep{2016A&A...592A.161N}. The earliest phase corresponds to $z\gtrsim2$. Although the assembly of galaxies has been confirmed as early as $z\sim6$ (e.g. \citealp{2014ApJ...792...15T}; \citealp{2019ApJ...883..142H}), galaxy overdensities that are bright in X-ray cannot be seen until $z\sim2$. This early phase of galaxy clusters is called a ``protocluster'' (\citealp{2016A&ARv..24...14O} and references therein). The star formation rate (SFR) density within protoclusters is predicted to increase monotonically {with decreasing redshift}, and about half of the stellar mass in the present clusters has formed by $z\sim2$ \citep{2017ApJ...844L..23C}. A redshift range of $1.3\lesssim z\lesssim2$ is known as a transition epoch, from the star-forming to the quiescent phase (e.g. \citealp{2013ApJ...779..138B}; \citealp{2014MNRAS.437..437A}). From $z\sim1.3$, as red-sequence galaxies are already present at the cluster core by $z\sim1$ {(\citealp{2004MNRAS.350.1005K}; \citealp{2012ApJ...746..188M})}, cluster galaxies are thought to evolve passively, via minor mergers, resulting in mass and size growth \citep{2007MNRAS.375....2D}. 

%P3:特に1.3<z<2に注目、environmental quenching 
This study focuses on the second-earliest phase, when a certain fraction of both actively star-forming galaxies and quenched galaxies coexist in a single cluster. Throughout this epoch, with decreasing redshift, both a rapid decrease in star formation activity, compared to field environments (\citealp{2013ApJ...779..138B}; \citealp{2019MNRAS.486.3047C}), and an increase in quenching efficiency \citep{2017MNRAS.465L.104N} are reported. However, the physical processes that are responsible for the star formation activity and environmental quenching are not well understood. One of the important quantities relating to star formation is the gas depletion timescale $\tau\equiv M_{\rm gas}/{\rm SFR}$, which requires information about both the molecular gas content and the star formation activity of galaxies. This timescale represents the effectiveness of star formation, where the shorter gas depletion timescale implies intense activity. Nearby ultraluminous infrared (IR) galaxies and distant submillimeter galaxies (SMGs) often have extremely short gas depletion timescales, and they are favorably explained by starbursts triggered by major mergers (\citealp{2010ApJ...714L.118D}; \citealp{2010ApJ...724..233E}; \citealp{2016ApJ...825..128L}).

%P4:high-z cluster galaxiesの分子ガス
For cluster galaxies at $z\gtrsim1.3$, submillimeter interferometers have discovered the ubiquity of a large amount of cold molecular gas ($M_{\rm mol \ gas}\gtrsim10^{10}M_{\odot}$), using mid-/low-$J$ transition lines of carbon monoxide (CO), the second most abundant molecule following molecular hydrogen ($\rm H_{2}$), or dust continuum emission for substantial samples 
(e.g. \citealp{2017A&A...608A..48D};
\citealp{2017ApJ...842...55L};
\citealp{2017ApJ...842L..21N};
\citealp{2017ApJ...844L..17W};
\citealp{2017ApJ...849...27R};  
\citealp{2018MNRAS.479..703C};
\citealp{2018Natur.556..469M};
\citealp{2018ApJ...856...72O};
\citealp{2018ApJ...867L..29W};  
\citealp{2019ApJ...872..117G};
\citealp{2019PASJ...71...40T};
\citealp{2019ApJ...887..183Z};
\citealp{2020A&A...641L...6D};
\citealp{2021ApJ...913..110C};
\citealp{2021A&A...652A..11J};
\citealp{2022ApJ...924...74A};
\citealp{2022ApJ...929...35W}).
These studies have been conducted by observing multiple sources with just a single telescope pointing, allowing an efficient survey of CO line emitters in dense cluster environments. This is in contrast to the observations of field galaxies, where multiple pointings are required to cover a large area of the sky and to build a statistically significant sample of galaxies {(e.g. \citealp{2016ApJ...833...67W}; \citealp{2019ApJ...882..138D})}. 

%P5:clusterとfieldの違い、high resolutionへの導入
The gas depletion timescale can be affected by various physical processes, such as mergers and large-scale gas inflow/outflow, as well as the overall environments. Close interactions and mergers of galaxies are often preferentially found in dense environments, and one will naturally predict a shorter gas depletion timescale in cluster galaxies. While several studies report a longer gas depletion timescale for cluster galaxies at high-redshift compared to field galaxies (e.g. \citealp{2017ApJ...842L..21N}; \citealp{2018ApJ...856..118H}; \citealp{2019PASJ...71...40T}; \citealp{2022ApJ...929...35W}), others find it to be consistent (e.g. \citealp{2019ApJ...872..117G}; \citealp{2019ApJ...887..183Z}; \citealp{2022ApJ...924...74A}) or even shorter \citep{2022ApJ...927..235A}. However, these studies are conducted using data with a relatively low angular resolution ($\gtrsim1''$), and they may just reflect the global properties averaged over the entire galaxy disk. For example, from their high-resolution observation of a $z=2.2$ SMG ALESS67.1 in the field environment, \cite{2017ApJ...846..108C} find a shorter gas depletion timescale at the central kiloparsec region than at the outskirts of the galaxy, presumably driven by a merger of two galaxies and the central gas inflow. On the other hand, cluster galaxies appear to harbor larger gas fractions than field galaxies (e.g. \citealp{2017ApJ...842L..21N}; \citealp{2018ApJ...856..118H}; \citealp{2019PASJ...71...40T}; \citealp{2022ApJ...929...35W}), which can result in an apparently longer gas depletion timescale for a fixed SFR. In addition, the evolutionary stages of galaxies can vary from cluster to cluster, even if they reside at the same redshift. Therefore, the true gas depletion timescale will likely vary throughout the galaxy disk and may also depend on the environment and evolutionary stage. A systematic high-resolution observation that probes the internal disk structures of galaxies in a number of clusters at different redshifts is necessary for a complete and proper understanding of the relationship between gas and star formation.

%P6:high-resolution studyの恩恵
Various sources of information that cannot be obtained in low-resolution observations become available via subarcsec-resolution observations at submillimeter wavelengths. First, subarcsec-resolution study probes the morphologies of dust-obscured star formation and the interstellar medium (ISM) of the galaxies, allowing us to characterize whether the dominant structure is smooth or clumpy throughout the galaxy. Second, the spatial extent, which is a key quantity for evaluating the structural evolution, can be measured. The spatial extent is also valuable for deriving the surface densities of the physical quantities that are conventionally used in the Kennicutt-Schmidt (KS) relation (\citealp{1959ApJ...129..243S}; \citealp{1989ApJ...344..685K}). Finally, multiple distinct components that {have been blended} in low-resolution images might be {resolved}. This allows us to study the effect of mergers on star formation and active galactic nucleus (AGN) activities. Up to the present, a considerable number of cluster galaxies at $z\gtrsim1.3$ have been studied in relatively low resolution. The important next step is to observe them in sufficiently high resolution, allowing detailed study of the spatial distribution and physical properties across the galaxy disks.

%P7:ALMAを用いた近年の銀河団研究
Over the past decade, the Atacama Large Millimeter/submillimeter Array (ALMA) has demonstrated its potential to observe distant galaxies with its unprecedented sensitivity and resolution \citep{2020RSOS....700556H}. In spite of the accessibility of CO lines in cluster galaxies, only one high-resolution study of CO lines has been made to date \citep{2019ApJ...870...56N}. Nevertheless, no high-resolution studies have been performed that contain information about both the molecular gas and the star formation properties toward high-redshift cluster galaxies.

%P8/9:本論の構成、宇宙論パラメータなど
In this paper, we report new observations of the resolved CO~$J$~=~2--1 line ($\nu_{\rm rest}=230.538$ GHz) emission, alongside the resolved 870 $\mu{\rm m}$ continuum emission, from 17 galaxies in XMMXCS J2215.9-1739 (hereafter, XCS J2215), a well-studied cluster at $z=1.46$. The aim of this paper is to measure the robust sizes of the CO~$J$~=~2--1 line and 870 $\mu{\rm m}$ continuum emissions and to obtain insights into the star formation properties and structural evolution of cluster galaxies. We first give a description of XCS J2215 and the ALMA observations in Section \ref{sec:Observations}. Our methods and main results are given in Section \ref{sec:Results}. In Section \ref{sec:Discussion}, we first investigate the resolved KS relation of the cluster galaxies and compare with the relation of the field galaxies. Then, we discuss the structural evolution of cluster galaxies on the basis of the spatial extent of the stellar, dust, and molecular gas components. Furthermore, we discuss the effects of galaxy mergers on the star formation properties and internal structures of galaxies. Finally, we summarize our conclusions in Section \ref{sec:Conclusions}. 

Throughout this paper, we assume a flat $\Lambda$ cold dark matter cosmology and adopt the cosmological parameters of $H_{0}=70 \ {\rm km \ s^{-1} \ Mpc^{-1}}$, $\Omega_{M}=0.3$, and $\Omega_{\Lambda}=0.7$. A redshift of $z=1.46$ corresponds to a cosmic age of 4.30 Gyr and gives a projected physical scale of 8.451 kpc/$''$. We use the Chabrier initial mass function \citep{2003PASP..115..763C} for the calculations of the stellar masses and SFRs.

\section{Observations} \label{sec:Observations}
\subsection{XMMXCS J2215.9-1738}\label{subsec:Subsec2.1}

\begin{figure*}[ht!]
\centering
\epsscale{1} 
\plotone{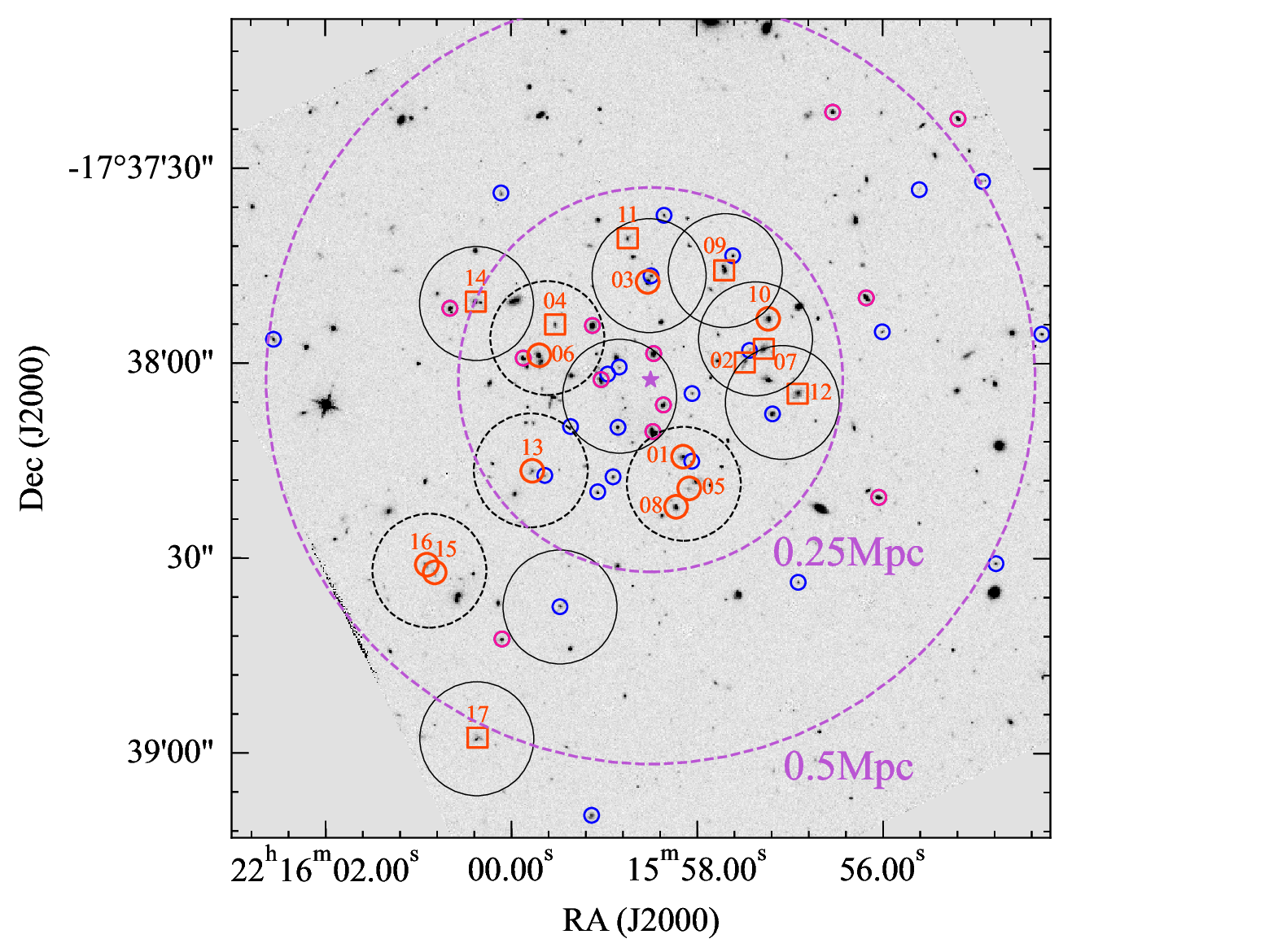}
\caption{The 2D distribution of 53 cluster galaxies in XCS J2215 on the HST WFC3/F160W-band image \citep{2017ApJ...846..120B}. The orange circles denote nine CO emitters with size measurements in both the CO~$J$~=~2--1 line and dust continuum emissions, while the orange squares denote the eight other emitters (Section \ref{subsec:Subsec3.2}). The numbers attached to these galaxies are the IDs defined in \cite{2017ApJ...841L..21H}. The blue and pink circles show the star-forming and passive galaxies classified by the UVJ diagram in \cite{2018ApJ...856..118H}. The purple star shows the cluster center determined by X-ray detection \citep{2006ApJ...646L..13S}, and two clustercentric radii of 0.25 Mpc and 0.5 Mpc are shown by the dashed purple circles. The solid and dashed black circles are the primary beams of the first (\#2012.1.00623.S) and the second (\#2017.1.00471.S) Band 7 observations, respectively.  \label{fig:Fig1}}
\end{figure*}

The cluster XCS J2215 at $z=1.46$ is detected as an extended X-ray-emitting source with a luminosity $L_{X}=2.9^{+0.2}_{-0.4}\times10^{14} \ {\rm erg \ s^{-1}}$ \citep{2010ApJ...718..133H} in the XMM Cluster Survey \citep{2001ApJ...547..594R}. Six cluster galaxies are initially spectroscopically confirmed in \cite{2006ApJ...646L..13S}.
A velocity dispersion obtained from 31 galaxies within $R_{\rm 200}=0.8\pm0.1$ Mpc \footnote{$R_{200}$ is the radius that encloses a density 200 times larger than the critical density at the same redshift.} is $\sigma_{v}=720\pm110 \ {\rm km \ s^{-1}}$ \citep{2010ApJ...718..133H}, corresponding to a cluster mass estimate of $M_{\rm cl}\sim3\times10^{14}M_{\rm\odot}$ under the assumption of a virialized halo. This implies that XCS J2215 is a mature cluster at $z\sim1.5$ and is likely to evolve into a Virgo-like cluster with a few times $10^{15}M_{\odot}$ at $z=0$ (\citealp{2013ApJ...779..127C}; \citealp{2014MNRAS.441L...1S}).

\cite{2011MNRAS.415.2670H} and \cite{2014MNRAS.439.2571H} present the spatial distribution of 639 [O{\sc ii}] line emitters at $z=$1.430--1.485 over a wide area ($42.5\times33.8 \ {\rm Mpc}^{2}$) around XCS J2215, obtained with the Suprime-Cam on the Subaru Telescope. An overdensity of [O{\sc ii}] emitters is confirmed within a central 3 Mpc region, suggesting that the cluster is in the star-forming phase, unlike similar massive clusters at lower redshifts. This is in line with the results from \cite{2015ApJ...806..257M}, who conducted SCUBA-2 850 $\mu$m and 450 $\mu$m observations to map the dust-obscured star-forming regions within a 0.8 Mpc region with an integrated SFR of $\sim1400M_{\rm\odot}{\rm yr^{-1}}$.

Recent ALMA follow-up studies have revealed that the galaxies found at the core of the cluster ($\lesssim0.6R_{200}$) are not only forming stars at a high rate, but also contain a large amount of cold gas, with an average value of $M_{\rm mol \ gas}=(6.2\pm2.6)\times10^{10} {M_{\rm \odot}}$ for galaxies detected in the CO~$J$~=~2--1 line (\citealp{2017ApJ...849..154S}; \citealp{2018ApJ...856..118H}). \cite{2017ApJ...841L..21H} report 17 of them (IDs: ALMA.01--ALMA.17) that have counterparts in optical or near-infrared (NIR) images. \cite{2018ApJ...856..118H} subsequently present the results of ALMA Band 7 observations and report detections of 870 $\mu$m continuum emission {with a mean flux $\langle f_{870\mu{\rm m}}\rangle=1.13\pm0.54$ mJy from eight cluster galaxies, considering a $4.6\sigma$ noise threshold.}
One of them, ALMA.18, is not observed in the CO~$J$~=~2--1 line, as it is located outside the field of view adopted in \cite{2017ApJ...841L..21H}. 

Figure \ref{fig:Fig1} shows the spatial distribution of 53 galaxies within a clustercentric radius of {$\sim0.5$ Mpc}. The cluster galaxy sample consists of [O{\sc ii}] emitters and spectroscopically confirmed galaxies \citep{2017ApJ...846..120B}. About 60\% of the members have optical morphologies that resemble ellipticals or S0s \citep{2009ApJ...697..436H}. We classify this sample into three groups: gas-rich members with CO~$J$~=~2--1 {line} detections (referred to as ``CO emitters'' hereafter), star-forming members, and passive members. In addition to the 17 CO emitters, 24 and 12 galaxies are classified as star-forming and passive members, respectively, based on the UVJ diagram \citep{2018ApJ...856..118H}. These 24 star-forming galaxies have no previous CO line detections and have not been targeted in ALMA observations. Six of the 12 passive galaxies are located within a clutercentric radius of 0.25 Mpc {($0.3R_{200}$)}, suggesting that the quenching of the star formation has already begun {at the cluster core.} On the other hand, CO emitters are found at the outskirts of the cluster core, where passive galaxies dominate (Figure \ref{fig:Fig1}). \cite{2017ApJ...841L..21H} find that CO emitters are distributed at the edge of the virialized region defined by \cite{2015MNRAS.448.1715J} in the phase-space diagram. This suggests that CO emitters have entered the cluster more recently than passive members and {are an important population} for understanding the transition from the star-forming to the passive phase. 

Overall, XCS J2215 is an unprecedentedly well-studied, high-redshift cluster with multiwavelength observations, which has notable star-forming properties at the cluster core. In this study, we select 17 cluster galaxies that have CO~$J$~=~2--1 line detections for our sample, adopting the same IDs as labeled in \cite{2017ApJ...841L..21H}. We do not include ALMA.18 in our sample, since the main purpose of this study is to discuss the spatial distribution of both the CO~$J$~=~2--1 line and dust continuum emissions near the cluster core.

\subsection{Stellar Mass, SFR, and Hubble Space Telescope Size} \label{subsec:Subsec2.2}

For 53 cluster galaxies, we derive the stellar mass by using the {\tt FAST} code \citep{2009ApJ...700..221K} to fit the spectral energy distribution (SED) generated from optical to mid-NIR photometry \citep{2018ApJ...856..118H}, then estimate the SFR using the ``SFR ladder'' \citep{2011ApJ...738..106W}. The IR luminosities are derived by scaling the average SED of the ALESS SMGs with $L_{\rm IR}<10^{11.9}L_{\odot}$ (\citealp{2015ApJ...806..110D}) to the observed 870 $\rm{\mu m}$ flux density, then integrating it from 8 to 1000 $\rm \mu m$. The total IR luminosities are then converted to the dust-obscured SFRs (SFR$\rm _{IR}$; \citealp{1998ARA&A..36..189K}) for galaxies that are detected in the 870 $\rm\mu m$ continuum, with an $8\sigma$ noise threshold (Section \ref{subsec:Subsec3.2}). We use Subaru/Suprime-Cam $R_{c}$-band ($\lambda_{\rm obs}=6517$ \AA) and $i'$-band ($\lambda_{\rm obs}=7671$ \AA) data, and compute the UV luminosities by $L_{2800}=4\pi d_{L}^{2}\nu_{2800} f_{2800}$, where $d_{L}$ is the luminosity distance and $\nu_{2800}$ is the frequency at 2800 \AA. We then use this luminosity to derive the unobscured SFRs (SFR$\rm_{UV}$; \citealp{2011ApJ...738..106W}). The derived total $\rm SFRs \ (= SFR_{IR}+SFR_{UV})$ are mostly consistent with $\rm SFR_{SED}$ \citep{2018ApJ...856..118H} within a 0.7~dex scatter. For galaxies without an 870 $\rm\mu m$ detection brighter than the $8\sigma$ noise level, we use the SFRs derived from the SED fitting \citep{2018ApJ...856..118H}. We note that the $\rm SFR_{SED}$ of these galaxies might be underestimated, since their 5--8$\sigma$ 870 $\mu$m continuum fluxes are not taken into account for the SED fitting. As we do not include these galaxies in the main analysis of this paper (Section \ref{subsec:Subsec3.2}), these SFRs do not affect the following discussion and conclusions.

The derived $\rm SFR_{IR}$ and $\rm SFR_{UV}$ are listed in the second and third columns of Table \ref{tab:physical properties}. The dust-obscured star formation ($\rm SFR_{IR}$) is dominant for all of the CO emitters that are detected in the 870 $\mu$m continuum, while the unobscured star formation ($\rm SFR_{UV}$) accounts for 1.7\%--10.9\% of the overall star formation. {\cite{2019ApJ...882..140B} report a median fraction of unobscured star formation of 8.5\% for 16 ASPECS CO emitters and two additional CO emitters. This implies that dust-obscured star formation dominates the star formation in CO emitters with $M_{\star}\gtrsim10^{10}M_{\odot}$, regardless of their environments (see also \citealp{2017MNRAS.466..861D}; \citealp{2017ApJ...850..208W}; \citealp{2022ApJ...928...88M}).}

\begin{figure}[t]
\centering
\epsscale{1.17}
\plotone{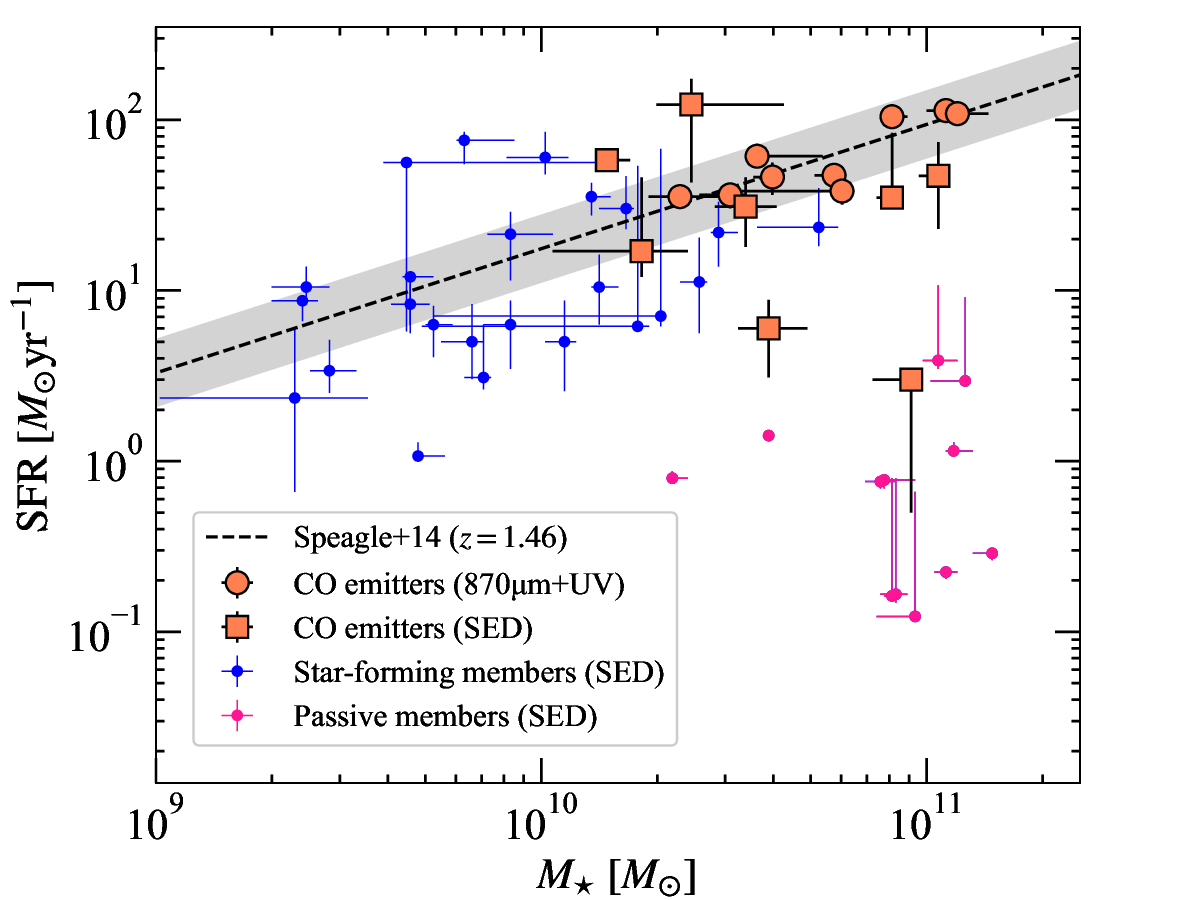}
\caption{The relation between stellar mass and SFR {of} 53 cluster members in XCS J2215. The symbol colors are the same as in Figure \ref{fig:Fig1}. For comparison, we show the star-forming MS at $z=1.46$ with the grey shaded region (with a $\pm0.2$ dex scatter; \citealp{2014ApJS..214...15S}).\label{fig:Fig2}}
\end{figure}

In Figure \ref{fig:Fig2}, we compare the stellar mass and the SFR of 53 cluster galaxies. CO emitters and passive members have comparable mass, but CO emitters are systematically located at the massive end of the star-forming main sequence (MS), at $z=1.46$ \citep{2014ApJS..214...15S}. The segregation between the CO emitters and the star-forming members in Figure \ref{fig:Fig2} is likely due to selection bias against the mass of the system.

Two of the 17 CO emitters (ALMA.11 and ALMA.14) host an AGN. These AGNs are identified based on their large flux ratio between the [N{\sc ii}] $\lambda$6584 line and the H$\alpha$ flux, both taken from Very Large Telescope/KMOS spectroscopy \citep{2019A&A...626A..14M}. ALMA.14 is also classified as an AGN by IR colors \citep{2010ApJ...718..133H}. None of the CO emitters are associated with an X-ray source brighter than $L_{X}\gtrsim 0.8\times10^{42} {\rm erg \ s^{-1}}$ \citep{2010ApJ...718..133H}, suggesting the minimal effect of AGNs {on galaxy properties} in these sources.

We also derive the half-light radius $R_{e,1.6{\rm\mu m}}$ (hereafter, 1.6 $\mu$m size) of the CO emitters from the F160W band on the Hubble Space Telescope (HST) Wide Field Camera 3 (WFC3), following the same procedure as presented in \cite{2016MNRAS.458.3181C}. We  assume a single S{\'{e}}rcic profile using GALFIT \citep{2010AJ....139.2097P}. The results of $R_{e,1.6{\rm\mu m}}$ are listed in the tenth column of Table \ref{tab:fitting}, providing an average size and standard deviation of $\langle R_{e,1.6{\rm\mu m}} \rangle=3.07\pm0.97$ kpc. We find that most of the galaxies have a S{\'{e}}rcic index around unity, indicating that the optical emission can nearly be characterized by an exponential disk profile.

\subsection{ALMA Observations} \label{subsec:Subsec2.3}
In this subsection, we describe the ALMA observations of {the} CO emitters in XCS J2215. {We use the Common Astronomy Software Application package ({\tt CASA}), version 5.1.1 \citep{2007ASPC..376..127M}, for the calibration of the ALMA data introduced below.}

\subsubsection{Band 3 Data}

The CO~$J$~=~2--1 line emission from the core of XCS J2215 is observed using ALMA Band 3 (Project ID: 2017.1.00471.S). Three {mosaic} pointings with a primary beam of FWHM$\sim58''$ cover all 17 CO emitters. The observations were carried out in 2017 December, with 42--49 antennas, as an ALMA Cycle 5 program. The range of the baseline length is 15.1--1397.8 m, corresponding to 11.6--1075.2 k$\lambda$ in the {\textit uv} plane. The maximum recoverable scale (MRS) ranges from $5\farcs4$ to $6\farcs4$, which is much larger than the typical angular size ($2''$--$3''$) of high-redshift galaxies, ensuring that all spatial structures are properly observed. The integration time for each pointing was $\sim4.3$ hr. Of the four spectral windows (SPWs) with central frequencies of 93, 94.7, 105, and 107 GHz, we used two adjacent SPWs on the lower sideband for the CO line observations. These SPWs cover the frequency range 92.06--95.64 GHz, enabling us to detect the CO~$J$~=~2--1 line emission ($\nu_{\rm rest}=230.538$ GHz) from $z=$1.41--1.50. This includes all of the possible redshifts probed by narrowband surveys of [O{\sc ii}] emitters \citep{2014MNRAS.439.2571H}. We use correlators in Frequency Division Mode, and each SPW consists of 480 channels over a bandwidth of 1.875 GHz, resulting in a spectral resolution of 3.9 MHz. This spectral resolution corresponds to $\Delta V= (3.9 \ {\rm MHz}/94 \ {\rm GHz})\times c\simeq12.4$~km~s$^{-1}$ (where $c$ is the speed of light $c\simeq 3\times10^{5}$~km~s$^{-1}$) in terms of the line-of-sight velocity. We average the spectral channels to 50~km~s$^{-1}$ bins and clean the cube down to $1.5\sigma$ by using {the} {\tt CASA/tclean} task. In order to obtain high-resolution maps, we adopt Briggs weighting with a robust parameter of +0.5, resulting in a synthesized beam size of $0\farcs46\times 0\farcs32$ ($3.89\times2.70$ kpc), with a position angle (PA) $=-79.1^{\circ}$, and an rms level of $\sim$75--80 $\mu{\rm Jy \ beam^{-1}}$ per 50~km~s$^{-1}$.

One of the main goals of this study is to spatially resolve the CO~$J$~=~2--1 line emission from high-redshift cluster galaxies and measure their effective radii. Nonetheless, previous ALMA Band 3 observations with the compact array configuration \citep{2017ApJ...841L..21H} are still useful for accurately measuring the total flux in the short {\textit uv} {range}. Therefore, to increase the sensitivity in the shorter {\textit uv} {range}, we combine the new extended array data with the previous compact data where 15.1--639.9 m (=11.6--492.2 k$\lambda$) are covered, using {\tt CASA/concat} task. With the combined visibility data, we make low-resolution maps with a robust parameter of +2.0, resulting in a synthesized beam size of $0\farcs64\times0\farcs48$ ($5.41\times4.06$ kpc), with a ${\rm PA}=-75.1^{\circ}$, and {an rms level of $\sim$65--68 $\mu{\rm Jy \ beam^{-1}}$ per 50~km~s$^{-1}$}. We use the spectra extracted with a $2\farcs0$ aperture in the low-resolution maps to define the velocity range of the CO~$J$~=~2--1 line emission (Appendix \ref{sec:CO (2-1)spectra}). We also evaluate the significance of the detection in the low-resolution maps. The measured signal-to-noise ratio (S/R) ranges from 7.1 to 33.5 in the velocity-integrated CO~$J$~=~2--1 maps (Table \ref{tab:fitting}).

\subsubsection{Band 7 Data}
We combined ALMA Band 7 data obtained from two separate projects (Project IDs: 2012.1.00623.S and 2017.1.00471.S) {to characterize the 870 $\rm\mu m$ continuum emission in all 17 CO emitters}. The first observation {was conducted in on 2015 July 19, using 39 12 m antennas and eight pointings. The second observation was conducted on 2018 September 22, using 44 12 m antennas and four pointings}. Each pointing has a primary beam of FWHM$\sim17''$, and we show the pointing positions for both observations in Figure \ref{fig:Fig1}. Four SPWs have central frequencies of 336.5, 338.4, 348.5, and 350.5 GHz, with 2 GHz bandwidth. We integrated all the SPWs for the analyses of {the} continuum emission. The ranges of baseline length and the integration times for each pointing are 15.1--1574.4 m ($=$17.3--1573.5 k$\lambda$) and 7 minutes for the first eight pointings and 15.1--1397.8 m ($=$17.3--1601.5 k$\lambda$) and 5 minutes for the second four pointings.

We use these visibility data for both imaging and visibility analyses. We create the deconvolved images by adopting natural weighting, with a robust parameter of +2.0. The synthesized beam sizes are $0\farcs18\times0\farcs16$ ($1.52\times1.35$ kpc), with a ${\rm PA} =45^{\circ}$ for the first eight pointings and $0\farcs24\times0\farcs22$ ($2.03\times1.86$ kpc), with a ${\rm PA} =56^{\circ}$ for the second four pointings, with rms levels of $\sim64$--$71 \ {\rm \mu Jy \ beam^{-1}}$ and $\sim48$--$51$ ${\rm \mu Jy \ beam^{-1}}$, respectively. Of the 17 CO emitters, we identify 15 and 10 detections of the 870 ${\mu {\rm m}}$ continuum emission at S/N $>$4.5 and S/N $>$8 (Table \ref{tab:fitting}).

\begin{figure*}[t]
\epsscale{1.17}
\plotone{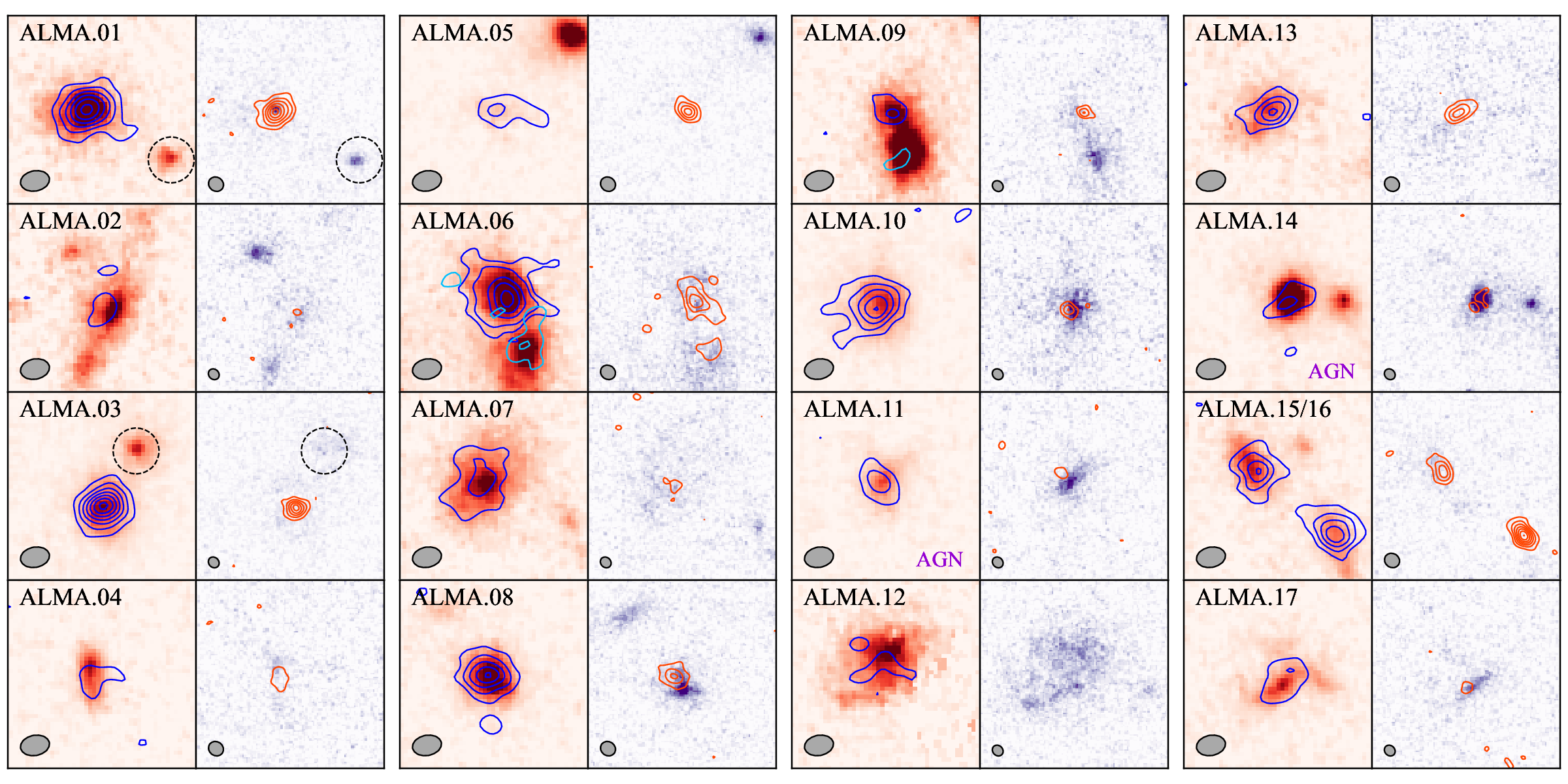}
\caption{The spatial distributions of the CO~$J$~=~2--1 line (blue contours) and 870 $\rm\mu m$ continuum emission (orange contours) of 17 CO emitters. The contours are drawn every 3$\sigma$. Each panel shows a region of {$\sim3''\times3''$ ($25$ kpc$\times25$ kpc in physical scale)}. We show the synthesized beams of the ALMA CO~$J$~=~2--1 line ($0\farcs46\times 0\farcs32$) and 870 $\rm\mu m$ continuum ($0\farcs18\times 0\farcs16$ or $0\farcs24\times0\farcs22$) data in the bottom left. The background images are the HST WFC3/F160-band (left) and ACS/F850LP-band (right) images. The dashed circles denote [O{\sc ii}] emitters at $z=$1.430--1.485. The cyan contours shown in the panels of ALMA.06 and ALMA.09 correspond to the CO~$J$~=~2--1 line emission from the companion.} \label{fig:Fig3}
\end{figure*}

\section{Analysis and Results} \label{sec:Results}
\subsection{Identification of Early-stage Mergers} \label{subsec:Subsec3.1}

In general, high-density environments {at high-redshift} contain a larger fraction of galaxy-galaxy mergers than the field, which can in turn significantly enhance the {star formation} activity in the gas-rich member galaxies involved {(e.g. \citealp{2018MNRAS.479..703C}).} We use the ALMA data along with the cluster member catalog to identify merger candidates among the 17 CO emitters found in XCS J2215. In Figure \ref{fig:Fig3}, we show the high-resolution contour maps of the CO~$J$~=~2--1 line and 870 $\mu$m continuum emissions of the 17 CO emitters. While the peak positions of the CO~$J$~=~2--1 line emission are mostly consistent with those of the optical counterparts, some galaxies show a spatial offset of $\sim0\farcs1$ between the 870 $\rm \mu m$ continuum emission and the images from the F850LP band on the HST Advanced Camera for Surveys (ACS; e.g., ALMA.08 and ALMA.11). We find a random offset of $\Delta {\rm R.A.}=\pm0\farcs05$ and $\Delta {\rm decl.}=\pm0\farcs03$, by comparing the source coordinates between the HST images and Gaia Early Data Release 3 \citep{2021A&A...649A...1G}, which can be attributed to the astrometric accuracy of the HST images. The spatial offset between the continuum and the HST images ($\sim0\farcs1$) is larger than this astrometric accuracy, suggesting that the offset is real and conceivably caused by dust extinction {\citep{2019MNRAS.488.1779C}}.

In this work, {we identify an early-stage merger if the CO emitters have a companion galaxy within a projected separation of 15 kpc ($\sim2''$).} ALMA.01 and ALMA.03 are associated with an [O{\sc ii}] emitter at separations of 13 kpc and 9 kpc, respectively. For ALMA.06 and ALMA.09, a close companion is not included in the cluster member catalog, but {we find a $6.4\sigma$ and a $3.7\sigma$ CO~$J$~=~2--1 line detection} with velocity separations of 1350~km~s$^{-1}$ and 1850~km~s$^{-1}$, respectively (Appendix \ref{sec:CO (2-1)spectra}). ALMA.15 and ALMA.16 consist of a close pair of CO emitters with a velocity separation of only 300~km~s$^{-1}$, suggesting that they are currently interacting with each other. ALMA.02, ALMA.05, ALMA.07, ALMA.08, and ALMA.14 also have close companions in the HST images, but these companions are neither cluster galaxies nor detected in the {CO lines}. Therefore, in summary, we find six CO emitters from the 17 candidates for early-stage mergers. Although we cannot rule out the possibility of projection effects, this indicates a higher merger fraction of 35\% (upper limit), compared to the fraction of coeval field galaxies (e.g. 11\% at $z=1.62$ in \citealp{2013ApJ...773..154L}). While the close proximity of the galaxies suggests gravitational interaction, clear tidal tails are undetected in both the HST and CO~$J$~=~2--1 images.

\begin{figure*}[t]
\epsscale{0.9}
\plotone{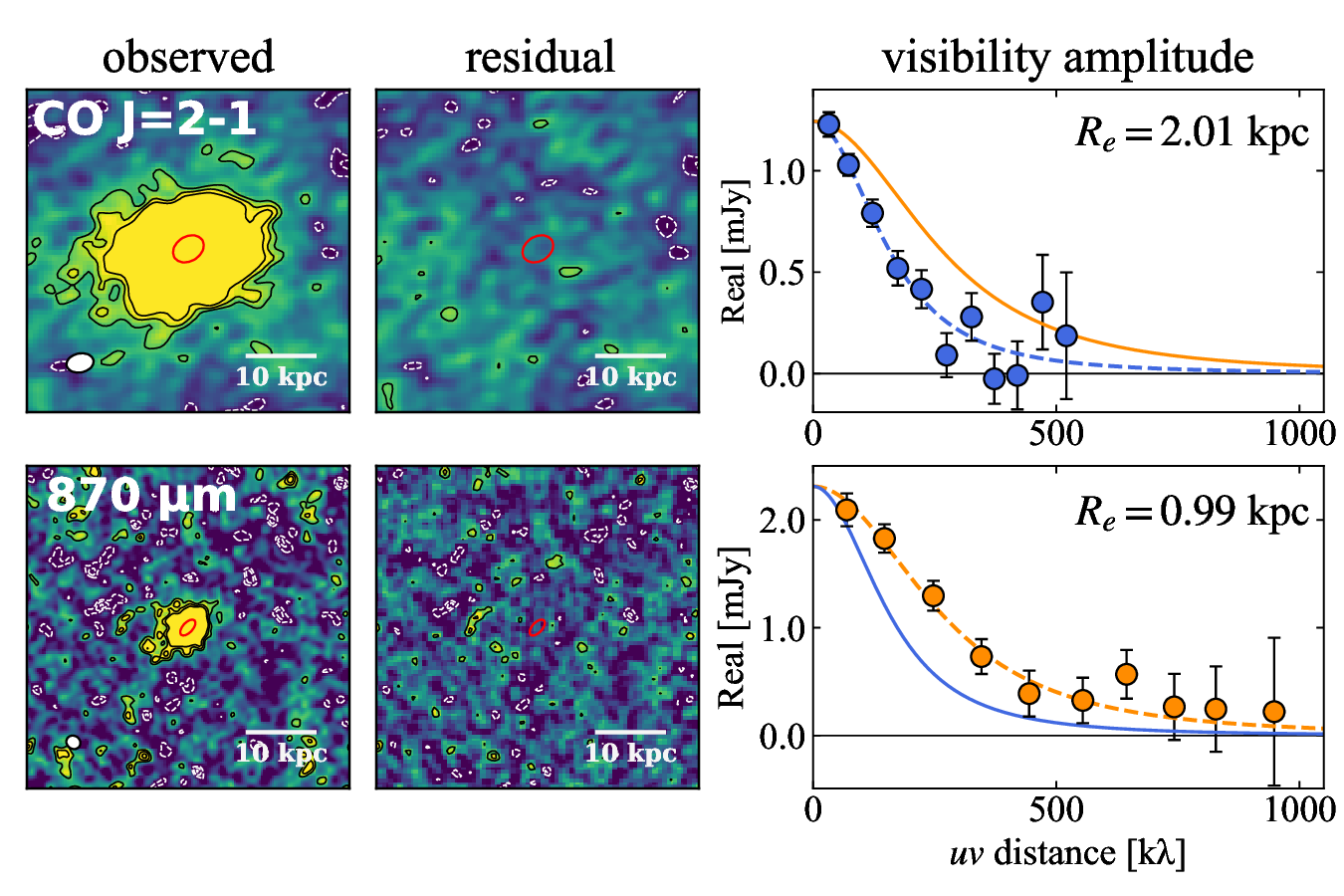}
\caption{A demonstration of {\textit uv}-plane analysis, showing an example from ALMA.01. Left: the observed dirty maps of the CO~$J$~=~2--1 line (top) and 870 $\rm\mu m$ continuum (bottom) emissions. The synthesized beam is shown in the bottom left. The effective radii of the best-fit model are shown as the red ellipses, in consideration of the axis ratio and PA. Middle: the dirty maps created from the visibility data, where the best-fit model of an exponential disk is subtracted. The contours of both the observed and the residual maps are $\pm2\sigma, \pm3\sigma$, and $\pm4\sigma$. Right: the real part of the visibility amplitude as a function of {\textit uv} distance with the best-fit model (dashed line). {As a comparison, the normalized profile of the other emission (solid line) is shown in both panels.}
The plots for the other galaxies are given in Appendix \ref{sec:uvamp}. 
\label{fig:Fig4}}
\end{figure*}

\subsection{Size Measurements on Visibility Plane} \label{subsec:Subsec3.2}

%P1
The high-resolution ALMA observations allow us to spatially resolve the CO~$J$~=~2--1 and 870 $\rm\mu m$ continuum emissions in the cluster galaxies at $z=1.46$ and measure effective radii that enclose a half of the total flux. A visual inspection suggests that the CO line emission is more extended than the 870 $\rm \mu m$ continuum emission in most of the CO emitters (Figure \ref{fig:Fig3}), but this apparent difference could be attributed to the synthesized beam, with a different resolution, and to varying S/Ns. It is, in principle, possible to obtain an effective radius by deconvolving an image with a clean beam when a source is detected at a sufficiently high S/N. However, even with ALMA, the S/Ns in submillimeter observations are {usually} much lower than those in optical and NIR observations, except for certain populations, such as optical dark galaxies (e.g. \citealp{2019ApJ...878...73Y}; \citealp{2019Natur.572..211W}; \citealp{2021MNRAS.502.3426S}). Moreover, image reconstruction for interferometer images depends on the {\textit uv} coverage of the visibilities taken from the observations and the CLEANing process. These factors occasionally make image-based analysis severely uncertain and may lead to erroneous interpretations. Therefore, we measure the effective radii of the CO~$J$~=~2--1 line and 870 $\rm \mu m$ continuum emissions directly from the visibility data, instead of using the {CLEANed} images {(e.g. \citealp{2015ApJ...799...81S}; \citealp{2019MNRAS.490.4956G};
\citealp{2020ApJ...900....1F}; \citealp{2020ApJ...901...74T})}.

%P2
With the ALMA Band 3 and Band 7 data, we perform visibility fittings in the {\textit uv} plane to obtain the best-fit model for the 17 CO emitters. The procedure for our visibility data analysis is as follows. In order to robustly measure an effective radius, we only use sources with S/N$>8$ in the low-resolution CO~$J$~=~2--1 or 870 $\rm\mu m$ continuum maps \citep{2019MNRAS.490.4956G}. This selection results in a sample of 15 for the CO~$J$~=~2--1 line emission and 10 for the 870 $\mu$m continuum emission. Recent {high-resolution ALMA} studies suggest that dusty star-forming galaxies at $z\sim$1--3 have an exponential profile, with a S{\'{e}}rcic index of $n\sim1$ in dust continuum emission, signifying the dominance of a disk-like morphology  (\citealp{2016ApJ...833..103H}; \citealp{2018ApJ...861....7F}). Accordingly, we assume that the galaxies in our sample have similar structures in both the CO~$J$~=~2--1 line and 870 $\mu$m continuum emissions, and we fit the visibility data with a 2D exponential disk model ($n=1$) by using {\tt UVMULTIFIT} \citep{2014A&A...563A.136M}. The observed visibility contains the amplitude and phase information of the other sources within the primary beam, especially when they are close to the target. Therefore, to obtain the parameters that are purely attributable to a single galaxy of interest, we subtract clean components of the other CO emitters in {\textit uv} plane. The primary beam correction is taken into account in the outcomes of the {\textit uv} fitting. 

%P3
We first set all of the parameters that have been incorporated in {\tt UVMULTIFIT} as free parameters: source position, total flux, effective radius along the major axis, major-to-minor axis ratio $q$, and PA. As the size parameter in {\tt UVMULTIFIT} is supplied with the FWHM, we convert it to an effective radius, as $R_{e}=1.21\times {\rm FWHM}$ for an exponential disk model. For some galaxies with relatively low S/N, the fittings do not converge correctly or the axis ratio becomes $q=0$ or $q=1$. Thus, we refit the data by fixing the axis ratio to $q=1$ to determine the other parameters for these galaxies. Because {the} main goal of this analysis is to derive the spatial extent of the emission, we use the circularized radius given by $R_{e}=R_{e,{\rm elliptical}}\times\sqrt{q}$ for a fair comparison.
In ALMA.09 only, the effective radius of the 870 $\rm \mu m$ continuum emission becomes zero within the uncertainty, indicating that the emission is not spatially resolved with the current data. We thus exclude this result from the following discussion. Consequently, we successfully measure the effective radii of the CO~$J$~=~2--1 line emission for 15 galaxies and the 870 $\mu$m continuum emission for nine galaxies.

%Table1
\begin{deluxetable*}{lccccccccl}[t]
\tablenum{1}
\tablecaption{Fitting results of  {the CO emitters} \label{tab:fitting}}
\tablewidth{18pt}
\tablehead{
\colhead{ID}
& \colhead{$\rm{S/N}_{870{\rm\mu m}}$}
& \colhead{$f_{870{\rm\mu m}}$}
& \colhead{{$q_{870{\rm\mu m}}$\tablenotemark{\small{\rm a}}}}
& \colhead{$R_{e,870{\rm\mu m}}$ \tablenotemark{\small{\rm b}}} 
& \colhead{$\rm{S/N_{CO}}$}
& \colhead{$I_{\rm CO}$}
& \colhead{{$q_{\rm CO}$\tablenotemark{\small{\rm a}}}}
& \colhead{$R_{e,{\rm CO}}$ \tablenotemark{\small{\rm b}}}
& \colhead{{$R_{e,1.6{\rm\mu m}}$}} \\
%& \colhead{merger \tablenotemark{\tiny{\rm a}}}\\
\colhead{} 
& \colhead{} 
& \colhead{(mJy)} 
& \colhead{}
& \colhead{(kpc)} 
& \colhead{}
& \colhead{(Jy~km~s$^{-1}$)}
& \colhead{}
& \colhead{(kpc)}
& \colhead{{(kpc)}}
%& \colhead{} 
}
\startdata
ALMA.01 & 19.2 & $2.31\pm0.21$ & $0.52\pm0.14$ & $0.99\pm0.20$ & 33.0 & $0.75\pm0.04$ & $0.75\pm0.17$ & $2.01\pm0.40$ & $3.73\pm0.44$ \\
ALMA.02 & -- & -- & -- &  -- & 7.1 & -- & -- & -- & $3.07\pm0.09$ \\
ALMA.03 & 16.7 & $2.52\pm0.29$ & $0.78\pm0.19$ & $0.84\pm0.17$ & 33.5 & $0.90\pm0.05$ & -- & $1.70\pm0.20$ & $2.72\pm0.03$ \\
ALMA.04 & 5.1 & -- & -- & -- & 9.0 & $0.18\pm0.04$ & --  & $2.00\pm0.88$ & $1.62\pm0.03$ \\
ALMA.05 & 13.6 & $0.78\pm0.11$ & -- & $0.31\pm0.14$ & 9.2 & $0.25\pm0.05$ & -- & $1.77\pm0.65$ & $3.32\pm0.21$ \\
ALMA.06 & 12.0 & $2.40\pm0.34$ & $0.42\pm0.12$ & $2.00\pm0.47$ & 29.1 & $1.10\pm0.06$ & $0.84\pm0.14$ & $2.83\pm0.41$ & $2.39\pm0.03$ \\
ALMA.07 & 5.1 & -- & -- & -- & 14.8 & $0.56\pm0.05$ & $0.79\pm0.22$ & $3.06\pm0.67$ & $4.14\pm0.04$ \\
ALMA.08 & 9.6 & $0.99\pm0.19$ & -- & $0.85\pm0.26$ & 23.7 & $0.48\pm0.05$ & $0.65\pm0.31$ & $1.60\pm0.52$ & $2.26\pm0.02$ \\
ALMA.09 & 9.4 & $0.59\pm0.12$ & -- & $0.06\pm0.40$ &  8.5 & $0.12\pm0.04$ & -- & $2.38\pm1.33$ & 
$3.14\pm0.09$ \tablenotemark{\small{\rm c}}  \\
ALMA.10 & 8.7 & $0.94\pm0.22$ & $0.72\pm0.52$ & $0.49\pm0.30$ & 26.4 & $0.68\pm0.04$ & $0.76\pm0.16$ & $2.32\pm0.42$ & $3.50\pm0.04$ \\
ALMA.11 & 5.5 & -- & -- & -- & 13.2 & $0.31\pm0.06$ & -- & $1.13\pm0.50$  & $3.42\pm0.12$ \\
ALMA.12 & -- & -- & -- & -- & 7.3 & -- & -- & -- & $4.21\pm0.06$ \\
ALMA.13 & 10.6 & $0.83\pm0.14$ & -- & $0.64\pm0.19$ & 18.8 & $0.45\pm0.06$ & -- & $1.29\pm0.41$ & $3.62\pm0.09$ \\
ALMA.14 & 4.5 & -- & -- & -- & 10.9 & $0.20\pm0.03$ & -- & $0.98\pm0.40$ & $1.20\pm0.02$ \\
ALMA.15 & 21.1 & $1.37\pm0.13$ & -- & $0.47\pm0.09$ & 17.9 & $0.39\pm0.04$ & -- & $1.17\pm0.28$ & $2.38\pm0.04$ \\
ALMA.16 & 11.4 & $0.79\pm0.13$ & -- & $0.55\pm0.18$ & 15.5 & $0.48\pm0.05$ & -- & $1.73\pm0.37$ & $2.45\pm0.04$ \\
ALMA.17 & 4.5 & -- & -- & -- & 10.6 & $0.34\pm0.05$ & $0.47\pm0.26$ & $2.10\pm0.87$ & $5.02\pm0.18$
\enddata
\tablenotetext{\small{\rm a}} {\ {The minor-to-major axis ratio. Values are only shown for the CO emitters that are successfully fitted in the {\textit uv} plane, setting $q$ as a free parameter.}}
\tablenotetext{\small{\rm b}} {\ {The circularized effective radii.}}
\tablenotetext{\small{\rm c}} {\ {The 1.6 $\mu$m size of the companion galaxy.}}
\end{deluxetable*}

%P4
In order to demonstrate the reliability of our fittings in the {\textit uv} plane, we show the dirty maps, the residual maps, and the real part of the visibility data as a function of the {\textit uv} distance {of ALMA.01} in Figure \ref{fig:Fig4}. The residual maps are created after subtracting the best-fit model in the {\textit uv} plane. In both emissions, the residual maps do not show any peaks above the $\sim3\sigma$ level around the {position of ALMA.01}, which implies that the observed emissions are well characterized by the best-fit model. The visibility amplitudes along the {\textit uv} distance underscore the reliability of the fittings. The descending curves of the visibility amplitude indicate the difference of the spatial extents in both emissions. The visibility amplitude holds constant along the {\textit uv} distance if the source is unresolved. In contrast, the amplitude drops sharply for an extended source. In Figure \ref{fig:Fig4}, the visibility amplitude of the CO~$J$~=~2--1 line emission drops at shorter {\textit uv} distances ($<500 {\rm k\lambda}$), while the visibility amplitude of the 870 $\rm\mu m$ continuum emission extends up to $\sim1000$ k$\rm\lambda$, indicating that the CO~$J$~=~2--1 line emission is more extended than the 870 $\rm\mu m$ continuum emission. The visibility amplitudes for the rest of the 14 CO emitters are provided in Appendix \ref{sec:uvamp}. 

\subsection{Size Comparison} \label{subsec:Subsec3.3}

\begin{figure}[t]
\epsscale{1.2}
\plotone{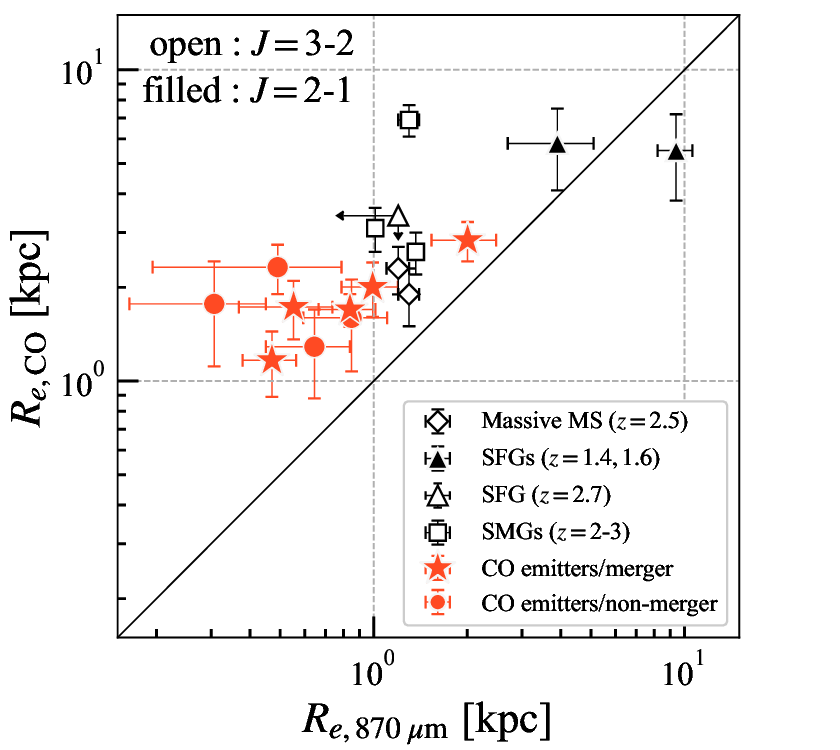}
\caption{Comparison of the effective radii of the CO~$J$~=~2--1 line and 870 $\mu$m continuum emissions. A black solid line corresponds to the equality of both radii. All of the nine CO emitters {(orange)} fall above the line, indicating that the cold gas component is more extended than the dust component. For comparison, galaxies where both radii are measured are shown (\citealp{2017ApJ...841L..25T}; \citealp{2017ApJ...846..108C}; \citealp{2018ApJ...863...56C}; \citealp{2019ApJ...876..112R}; \citealp{2020ApJ...899...37K}). For these galaxies, the cold molecular gas is traced by either the CO~$J$~=~3--2 line (open markers) or the CO~$J$~=~2--1 line (filled markers). \label{fig:Fig5}}
\end{figure}

The averages and standard deviations of the effective radii in each emission are $\langle R_{e,{\rm CO}}\rangle=1.87\pm0.59$ kpc and $\langle R_{e,870{\rm\mu m}}\rangle=0.79\pm0.47$ kpc. For the nine CO emitters for which both effective radii are available, we obtain $\langle R_{e,{\rm CO}}\rangle=1.82\pm0.48$ kpc. {Recent TNG50 simulations have demonstrated that the effective radii of the dust continuum emission are consistent with the effective radii of the dust-obscured star formation, and the radii of the dust continuum emission are similar within the variation of $\sim0.1$ dex across the far-IR wavelengths (\citealp{2022MNRAS.510.3321P}; see also \citealp{2019MNRAS.488.1779C}). Therefore, we adopt the 870 $\mu$m radii as quantities that trace the spatial extent of the dust-obscured star formation, as well as the distribution of the dust heated by UV radiation from the young stellar population. Figure \ref{fig:Fig5} shows a comparison of the spatial extents of dust-obscured star formation and cold molecular gas,} represented by the 870 $\mu$m continuum and CO~$J$~=~2--1 line emissions, respectively. We also show a compilation of high-redshift galaxies from the literature. We only include sources where both sizes are measured by ALMA observations: three SMGs at $z=$2--3 (\citealp{2017ApJ...846..108C}, \citealp{2018ApJ...863...56C}, \citealp{2019ApJ...876..112R}), two massive MS galaxies \citep{2017ApJ...841L..25T}, and three UV-extended MS galaxies at $z\sim2$ \citep{2020ApJ...899...37K}. Here, we only refer to the studies that trace cold molecular gas by a CO line of either $J=$3--2 or $J=$2--1.

As seen in Figure \ref{fig:Fig5}, all of the nine CO emitters lie above the line where the two effective radii are the same (the solid black line). The average ratio with standard deviation is $\langle R_{e,{\rm CO}}/R_{e,870{\rm\mu m}}\rangle=2.8\pm1.4$. When the individual measurement errors of the CO and 870 $\mu$m sizes are propagated, the average size ratio is larger than unity at the significance level of 12.4$\sigma$, which corroborates the cold-gas component being more extended than the dust-obscured star formation. This also holds true for almost all of the galaxies from the literature, with one exception of a UV-extended MS galaxy ALPS.2 at $z=1.55$ \citep{2020ApJ...899...37K}. By comparing with the 1.6 $\mu$m sizes, {we obtain the average ratios with standard deviation of $\langle R_{e, {\rm 1.6\mu m}}/R_{e, {\rm CO}}\rangle=1.7\pm0.6$ and $\langle R_{e, {\rm 1.6\mu m}}/R_{e, {\rm 870\mu m}}\rangle=4.9\pm2.7$}. We exclude ALMA.09 from this calculation, since the 1.6 $\mu$m size of ALMA.09 traces the companion galaxy instead of the CO emitter (Figure \ref{fig:Fig3}).

Figure \ref{fig:Fig5} indicates that the CO emitters in XCS J2215, except for ALMA.06, are smaller in both the CO~$J$~=~2--1 line and 870 $\rm\mu m$ continuum emissions than other star-forming galaxies. However, the star-forming properties of the adopted literature sample may be biased toward starburst outliers, like SMGs, or the extended outliers with a normal star-forming regime. Therefore, a control sample that consists of field galaxies with similar star-forming properties is necessary for a proper comparison. We note that similar {high-resolution} observations targeting $z\gtrsim1$ field galaxies have not been conducted to date. 

{A high-resolution observation of the cluster galaxies in SpARCS J0225 ($z=1.60$) is reported by \cite{2019ApJ...870...56N}, measuring deconvolved CO sizes from the ALMA images for eight CO emitters}. {They assume a 2D Gaussian profile and obtain the effective radii of $R_{e,{\rm CO}}=$2.5--6.4 kpc for seven galaxies, which are entirely larger than our measurements of $R_{e,{\rm CO}}$. In order to fairly compare the effective radii of the SpARCS J0225 galaxies to our sample, we conduct size measurements using the ALMA archival data (Project ID: 2017.1.01228.S) of the SpARCS J0225 observation. The procedure for the size measurement is the same as that described in Section \ref{subsec:Subsec3.2}. We obtain the effective radius of the CO~$J$~=~2--1 line emission for six out of the eight galaxies, where all of them show smaller values {by a factor of $1.7\pm0.7$} compared to the ones based on the deconvolved images. However, we find that three galaxies (J0225-281, J0225-371, and J0225-541) are still larger than 3 kpc (3.09, 4.60, and 5.36 kpc, respectively). More details of the fitting results for the SpARCS J0225 members can be found in Appendix \ref{sec:HST_CO}.}

{The size discrepancy of the CO emitters between XCS J2215 and SpARCS J0225 can be attributed to cluster-to-cluster variations, which possibly reflect the evolutionary stage. Nevertheless, the current data is insufficient to draw a definitive conclusion, since the information about SpARCS J0225, such as the cluster mass and the 2D distribution of the other members, is still limited. In addition, no other cluster samples at similar redshifts are available for detailed comparisons. Future ALMA observations of the resolved CO line, targeting well-studied clusters at equivalent redshift, are necessary.}

\subsection{Physical Properties}\label{subsec:Subsec3.4}
\subsubsection{Molecular Gas Masses}

We convert the total flux of the CO~$J$~=~2--1 line emission derived from the {\textit uv} fittings (Section \ref{subsec:Subsec3.2}) into the molecular gas mass $M_{\rm mol \ gas}$. In this work, we adopt {a luminosity ratio} of $L'_{{\rm CO}~J=1-0}/L'_{{\rm CO}~J=2-1}=1.2$, as a typical value of normal star forming galaxies (e.g. \citealp{2008AJ....136.2782L}), and a galactic CO-to-$\rm H_{2}$ conversion factor of $\alpha_{\rm CO}=4.36M_{\rm\odot}$ $\rm(K \ km \ s^{-1} \ pc^{-2})^{-1}$ (e.g. \citealp{2013ARA&A..51..207B}), {which includes} a 36\% correction of helium abundance (\citealp{2001ApJ...547..792D}).
The calculated molecular gas masses are listed in the fourth column of Table \ref{tab:physical properties}. {We find that the molecular gas masses derived from the {\textit uv} fittings are largely consistent with the ones derived from image-based analysis in \cite{2018ApJ...856..118H}, with a scatter of $\sim$0.1 dex.} 

% Table2
\begin{deluxetable*}{lccrrcccc}[t]
\tablenum{2}
\tablecaption{Physical properties of {the CO emitters} \label{tab:physical properties}}
\tablewidth{20pt}
\tablehead{
\colhead{}
& \multicolumn{4}{c}{The Integrated Properties}
& \multicolumn{2}{c}{The Central Region}
& \multicolumn{2}{c}{The Extended Region} \\
\colhead{}
& \multicolumn{4}{c}{}
& \multicolumn{2}{c}{($0<r<R_{e,870{\rm\mu m}}$)}
& \multicolumn{2}{c}{($R_{e,870{\rm\mu m}}<r<R_{e,{\rm CO}}$)} \\
\cline{2-9}
\colhead{ID}
& \colhead{$\rm SFR_{\rm IR}$}
& \colhead{$\rm SFR_{\rm UV}$}
& \colhead{$M_{\rm mol \ gas}$}
& \colhead{$M_{\rm \star}$ \tablenotemark{\small{\rm a}}}
& \colhead{$\log{\Sigma_{\rm SFR}}$} 
& \colhead{$\log{\Sigma_{\rm mol \ gas}}$}
& \colhead{$\log{\Sigma_{\rm SFR}}$} 
& \colhead{$\log{\Sigma_{\rm mol \ gas}}$} \\
\colhead{} 
& \colhead{($M_{\odot}{\rm yr}^{-1}$)}
& \colhead{($M_{\odot}{\rm yr}^{-1}$)}
& \colhead{($10^{10}M_{\odot}$)}
& \colhead{($10^{10}M_{\odot}$)}
& \colhead{($M_{\odot}{\rm yr^{-1}kpc^{-2}}$)} 
& \colhead{($M_{\odot}{\rm pc^{-2}}$)} 
& \colhead{($M_{\odot}{\rm yr^{-1}kpc^{-2}}$)}
& \colhead{($M_{\odot}{\rm pc^{-2}}$)}
}
\startdata
ALMA.01 & $102\pm9$ & $1.9\pm0.1$ & $10.8\pm0.6$ & $8.1^{+0.8}_{-0.5}$ & $1.22\pm0.18$ & $3.85^{+0.24}_{-0.21}$ & $0.58^{+0.28}_{-0.30}$ & $3.53^{+0.25}_{-0.25}$ \\
ALMA.03 & $111\pm13$ & $1.9\pm0.1$ & $13.0\pm0.8$ & $11.2^{+0.3}_{-1.2}$ & $1.40\pm0.19$ & $4.07^{+0.20}_{-0.19}$ & $0.76^{+0.21}_{-0.22}$ & $3.70^{+0.16}_{-0.16}$ \\
ALMA.04 & -- & $0.8\pm0.1$ & $2.6\pm0.6$ & $3.9^{+1.0}_{-0.7}$ & -- & -- & -- & -- \\
ALMA.05 & $35\pm5$ & $0.8\pm0.1$ & $3.7\pm0.7$ & $2.3^{+0.7}_{-0.4}$ & $1.77\pm0.41$ & $3.64^{+0.68}_{-0.45}$& $0.11^{+0.38}_{-0.37}$ & $3.25^{+0.38}_{-0.36}$ \\
ALMA.06 & $106\pm15$ & $2.3\pm0.1$ & $15.8\pm0.9$ & $12.0^{+2.4}_{-0.8}$ & $0.62\pm0.21$ & $3.62^{+0.23}_{-0.22}$ & $0.47^{+0.37}_{-0.39}$ & $3.32^{+0.33}_{-0.33}$ \\
ALMA.07 & -- & $2.8\pm0.1$ & $8.1\pm0.7$ & $8.1^{+0.2}_{-0.7}$ & -- & -- & -- & -- \\
ALMA.08 & $44\pm8$ & $3.7\pm0.1$ & $6.9\pm0.7$ & $5.8^{+0.3}_{-0.6}$ & $0.98\pm0.27$ & $3.83^{+0.40}_{-0.30}$ & $0.42^{+0.45}_{-0.46}$ & $3.51^{+0.42}_{-0.42}$ \\
ALMA.09 & $26\pm5$ & $3.0\pm0.1$ & $1.8\pm0.6$ & $10.7^{+0.3}_{-1.2}$ & -- & -- & -- & -- \\
ALMA.10 & $42\pm10$ & $4.6\pm0.1$ & $9.8\pm0.6$ & $3.4^{+0.3}_{-0.4}$ & $1.44\pm0.53$ & $3.81^{+0.56}_{-0.54}$ & $-0.05^{+0.28}_{-0.28}$ & $3.44^{+0.19}_{-0.18}$ \\
ALMA.11 & -- & $4.2\pm0.1$ & $4.5\pm0.8$ & $1.8^{+0.6}_{-0.8}$ & -- & -- & -- & -- \\
ALMA.13 & $37\pm6$ & $1.5\pm0.1$ & $6.5\pm0.9$ & $6.0^{+0.4}_{-2.7}$ & $1.15\pm0.27$ & $4.01^{+0.40}_{-0.31}$ & $0.51^{+0.43}_{-0.46}$ & $3.68^{+0.39}_{-0.39}$ \\
ALMA.14 & -- & $4.0\pm0.1$ & $2.8\pm0.4$ & $9.1^{+0.0}_{-1.9}$ & -- & -- & -- & -- \\
ALMA.15 & $61\pm6$ & $0.7\pm0.1$ & $5.7\pm0.6$ & $3.6^{+1.7}_{-0.2}$ & $1.64\pm0.17$ & $4.08^{+0.28}_{-0.22}$& $0.77^{+0.30}_{-0.32}$ & $3.74^{+0.27}_{-0.26}$ \\
ALMA.16 & $35\pm6$ & $1.4\pm0.1$ & $6.9\pm0.8$ & $3.1^{+0.3}_{-0.5}$ & $1.26\pm0.30$ & $3.86^{+0.36}_{-0.32}$ & $0.16^{+0.28}_{-0.30}$ & $3.51^{+0.23}_{-0.23}$ \\
ALMA.17 & -- & $2.5\pm0.1$ & $4.9\pm0.7$ & $2.5^{+1.8}_{-0.5}$ & -- & -- & -- & -- \\
\enddata
\tablenotetext{\small{\rm a}}{~Taken from \cite{2018ApJ...856..118H}.}
\end{deluxetable*}

\subsubsection{Surface Densities of SFR and $M_{mol \ gas}$}
For nine CO emitters, we have measured the effective radii of both the CO~$J$~=~2--1 line and 870 ${\mu}$m continuum emissions. This enables us to estimate the surface densities of the SFR ($\Sigma_{\rm SFR}$) and molecular gas mass ($\Sigma_{\rm mol \ gas}$) for the area enclosed by these radii. The derivation of the effective radii can be difficult using low-resolution data alone, and {thus} the optical sizes of the galaxies are often used for deriving the surface densities {of the SFR and molecular gas mass}. One caveat is that the spatial extent could be different {between the stellar component and the dust or cold-gas components, as demonstrated in Section \ref{subsec:Subsec3.3}.} To obtain better constraints on these quantities, we utilize the best-fit model and the effective radii of nine CO emitters.

We define two regions, with different scales, for each galaxy: the ``central'' region, which is defined as the area enclosed within the effective radius of the 870 $\rm\mu m$ continuum ($0<r<R_{e,870{\rm\mu m}}$), and the ``extended'' region, which is defined as the area enclosed between the effective radii of the 870 $\mu$m continuum and CO~$J$~=~2--1 line emissions ($R_{e,870{\rm\mu m}}<r<R_{e,{\rm CO}}$). Assuming a 2D exponential disk model, we calculate the fraction of the enclosed flux {$A$} between the specific radii {($R_{1}$ and $R_{2}$)} as
\begin{align}
{A}=\frac{\int_{R_{1}}^{R_{2}} \exp[-1.68r/R_{e}]r dr}{\int_0^{\infty} \exp[-1.68r/R_{e}]r dr} \ \ {(0\leq A\leq1)}
\end{align}

\noindent
so that the surface densities of the SFR and molecular gas mass can be written as 
\begin{align}
\Sigma_{\rm SFR}&=\cfrac{{A}_{870{\rm\mu m}}\times {\rm SFR}}{\pi (R_{2}^{2}-R_{1}^{2})} \\
\Sigma_{\rm mol \ gas}&=\cfrac{{A}_{\rm CO}\times M_{\rm mol \ gas}}{\pi (R_{2}^{2}-R_{1}^{2})}
\end{align}

\noindent
By definition, the fraction of the 870 $\rm\mu m$ flux density in the central region within $R_{e,870{\rm\mu m}}$ is ${A}_{870{\rm\mu m}}=0.5$. To evaluate the values and uncertainties of ${A}_{870{\rm\mu m}}$ of the extended region and ${A}_{\rm CO}$ of both the central and extended regions, we generate 10,000 exponential models, by adding the random error to the effective radius of the best-fit model. We regard the 16th--84th percentile range of 10,000 numerical values as the uncertainties. We then derive the errors on the surface densities of the SFR and molecular gas mass by the propagation of the errors. 

The calculated surface densities of the SFR and molecular gas mass for both the central and extended regions are listed in the sixth to ninth columns of Table \ref{tab:physical properties}. We obtain average SFR surface densities of $\langle\Sigma_{\rm SFR}\rangle=1.28\pm0.32$, $0.41\pm 0.27$ $M_{\odot}{\rm yr^{-1}kpc^{-2}}$ and average molecular gas mass surface densities of $\langle\Sigma_{\rm mol \ gas}\rangle=3.86\pm 0.16$, $3.52\pm 0.16$ $M_{\odot}{\rm pc^{-2}}$ for the central and the extended region, respectively. The uncertainties correspond to the standard deviations.

\section{Discussion} \label{sec:Discussion}

\subsection{Spatially Resolved View of Star Formation Regime }\label{subsec:Subsec4.1}

%KS則とは何か
One of the practical relations that encapsulates the information about the star formation properties is commonly referred to as the KS relation (\citealp{1959ApJ...129..243S}; \citealp{1989ApJ...344..685K}). This empirical relation illustrates the correlation between the surface densities of the gas mass and SFR, and can be expressed by the formula $\Sigma_{\rm SFR}\propto\Sigma_{\rm mol \ gas}^{N}$. The slope  of $N=1$ in this relation corresponds to a constant star formation efficiency (${\rm SFE}\equiv \Sigma_{\rm SFR}/\Sigma_{\rm mol \ gas}$) or a constant gas depletion timescale $\tau\equiv 1/{\rm SFE}$, providing a straightforward signpost of the remaining time until galaxies exhaust their molecular gas content.

The KS relations of nearby galaxies have been investigated on both a global scale (e.g. \citealp{1998ApJ...498..541K}; \citealp{2017ApJS..233...22S}) and a sub-kpc scale (e.g. \citealp{2008AJ....136.2846B}; \citealp{2010ApJ...722L.127O}; \citealp{2011ApJ...730L..13B}). 
%high-zのKS則について
On the other hand, because it is observationally expensive to address the distribution of cold molecular gas via CO lines at high-redshift, and therefore difficult to measure the size of the molecular gas component, only a handful of $z\gtrsim1$ galaxies have been studied in detail in the context of the KS relation. Some studies focus on normal star-forming galaxies (\citealp{2010ApJ...714L.118D}; \citealp{2010MNRAS.407.2091G}), and others study extreme sources that are bright and extended, such as SMGs (e.g. \citealp{2015ApJ...798L..18H}; \citealp{2017ApJ...846..108C}). The former studies claim the ``bimodality'' of the star formation regime at high redshift, a normal regime in star-forming disks, and a starburst regime induced by mergers. The surface densities derived for the central and extended regions (Section \ref{subsec:Subsec3.4}) allow us to probe the characteristics of high-redshift cluster galaxies in {terms of} the KS relation for the first time. 

\begin{figure*}[ht!]
\epsscale{1.1}
\plotone{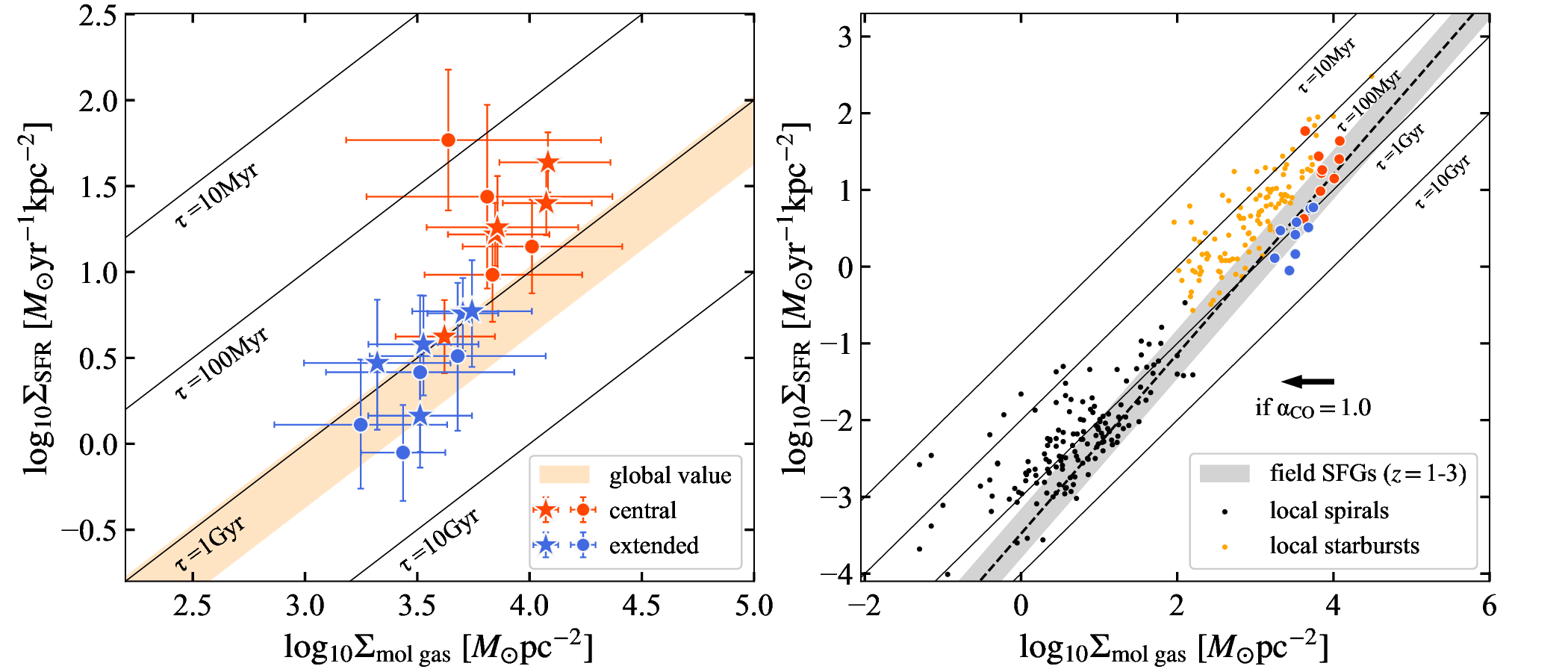}
\caption{The resolved KS relation of the nine CO emitters in XCS J2215. Left: a close-up view of the KS relation of the nine CO emitters. The red and blue plots correspond to the central and extended regions, respectively. The beige shaded region indicates the range of the global value of the gas depletion timescale among these galaxies. Mergers and nonmergers are discriminated by symbols, with stars indicating mergers and circles indicating nonmergers. Right: a comparison with the other galaxy populations from the literature, showing the wider range of surface densities in both axes. The black dots and orange dots are local spirals and local starbursts (\citealp{2019ApJ...872...16D}; \citealp{2021ApJ...908...61K}), respectively, and the black dashed line with the shaded region is a slope with $\pm0.32$ dex for normal star-forming galaxies at $z=$1--3 \citep{2010MNRAS.407.2091G}. The galactic value of $\alpha_{\rm CO}$ is adopted for all {of} the galaxies shown in this figure. The leftward arrow indicates the 0.64~dex shift of $\Sigma_{\rm mol gas}$ if $\alpha_{\rm CO}=1.0$ is adopted. \label{fig:Fig6}}
\end{figure*}

%Fig.6(左)に関する議論
The resolved KS relation for the nine CO emitters is shown in Figure \ref{fig:Fig6} (left). The shaded region in beige indicates the range of global values of gas depletion timescale, which is derived from the integrated quantities of each galaxy; $\tau_{\rm global}\equiv M_{\rm mol \ gas}/{\rm SFR}=0.9$--$2.4$ Gyr. The average and standard deviations are $\langle\tau_{\rm global}\rangle=1.49\pm0.46$ Gyr. Compared to the global values, the extended regions have a slightly shorter gas depletion timescale of $\langle \tau_{\rm ext} \rangle=1.42\pm0.73$ Gyr on average and the central regions have an even shorter timescale of $\langle \tau_{\rm cen} \rangle=0.48\pm0.27$ Gyr. {The shorter gas depletion timescale for the central region is consistent with the compactness of 870 $\mu$m continuum emission compared to CO~$J$~=~2--1 line emission at 12.4$\sigma$ significance (Section \ref{subsec:Subsec3.3}).} 

The central region tends to have a larger surface density than the extended region in both molecular gas mass and SFR. If the rates of increase are similar in both surface densities, then the gas depletion timescale should remain constant. Therefore, the decline of the gas depletion timescale at the central region {can be attributed to} higher SFR surface densities. {This finding, along with the evidence of compact dust emission (Section \ref{subsec:Subsec3.3}), indicates enhanced star formation activity in the central parts of the gas-rich cluster galaxies.}

%Fig.6(右)に関する議論
In Figure \ref{fig:Fig6} {(right), we compile the molecular gas mass and SFR surface densities of the nearby to high-redshift galaxies available in the literature, and compare them with the nine CO emitters in XCS J2215} (\citealp{2010MNRAS.407.2091G}; \citealp{2019ApJ...872...16D}; \citealp{2021ApJ...908...61K}). Compared to the local spirals, the CO emitters have higher surface densities in both SFR and molecular gas mass, by several orders of magnitude. Overall, the surface densities of the CO emitters are similar to the upper ends of the local starbursts. In addition, we find {that} both the central and extended regions {are consistent with the KS relation derived} from normal star-forming galaxies at $z=$1--3 \citep{2010MNRAS.407.2091G}. This suggests comparable depletion timescales {between cluster and field galaxies}, except for one galaxy, ALMA.05, which has the shortest gas depletion timescale in the central region of the nine CO emitters. {We conclude} that the star-forming properties of most of the cluster galaxies are similar to the ones of coeval field galaxies. 

A shorter gas depletion timescale {in} the central region is {similarly seen in} the $z=4.05$ SMG GN20 \citep{2015ApJ...798L..18H} and in the $z=2.12$ SMG ALESS67.1 \citep{2017ApJ...846..108C}. While the sample size is still limited, we tentatively suggest a general trend: that high-redshift star-forming galaxies have a shorter gas depletion timescale in the central region. This is not seen in relatively quiescent nearby galaxies, which follow a single slope in the KS relation (e.g. \citealp{2013AJ....146...19L}; \citealp{2021MNRAS.501.4777E}). On the contrary, \cite{2022A&A...659A.102S} find a shorter gas depletion timescale in the central regions of four (out of 16) nearby luminous IR galaxies (LIRGs). Our results are consistent with this, although the data for our cluster galaxies are limited, both in sample size and angular resolution.

\begin{figure*}[ht]
\epsscale{1.1}
\plotone{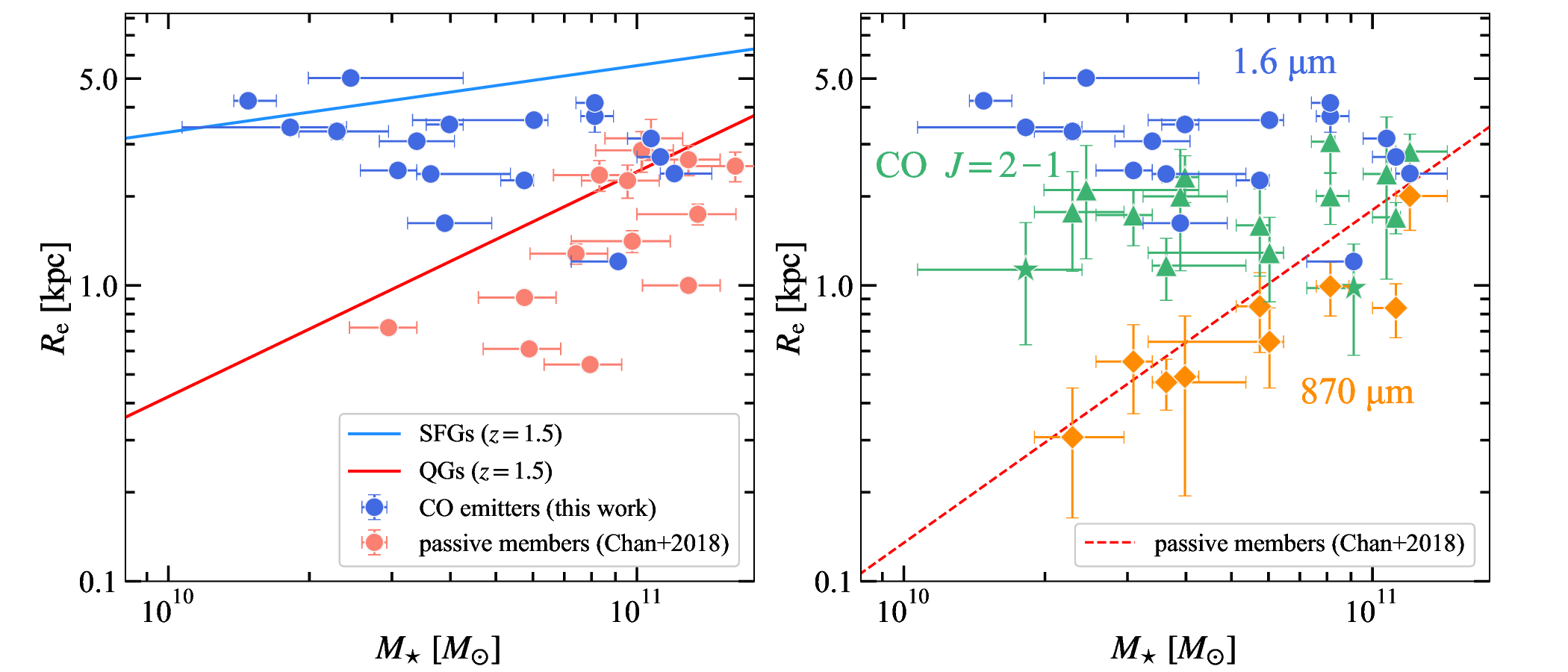}
\caption{{Stellar mass-size distribution of the galaxies in XCS J2215. 
Left: HST/1.6 $\mu$m sizes are shown for 17 CO emitters (blue circles) and 14 spectroscopically confirmed passive members (red circles; \citealp{2018ApJ...856....8C}). The solid lines correspond to the best-fit mass-size relation of star-forming (blue) and passive (red) galaxies at $z=1.5$ \citep{2014ApJ...788...28V}. 
Right: comparison of the sizes of the CO emitters measured from different tracers. The blue circles, green triangles or stars, and orange diamonds markers indicate the effective radii of the HST/1.6 $\mu$m, CO~$J$~=~2--1 line and 870 $\mu$m continuum, respectively. Two AGNs (ALMA.11 and ALMA.14) are shown with green stars for the CO size. The red dashed line is the best-fit mass-size relation of the passive members of XCS J2215 at 1.6 $\mu$m, as presented in the left panel.} \label{fig:Fig7}}
\end{figure*}

%留意事項1:表面密度の計算にどの半径を使うか。
{One caveat in this comparison} (Figure \ref{fig:Fig6} right) is that the past studies adopt different regions defined by other tracers when {the CO or 870 $\mu$m sizes} are not available. \cite{2010MNRAS.407.2091G} use the average effective radii of multiple indicators, such as the H$\alpha$ line and optical/UV stellar light. \cite{2019ApJ...872...16D} and \cite{2021ApJ...908...61K} {use the} diameter that contains $\sim95$\% of the H$\alpha$ flux. In {our case, the SFR surface densities of the central region would be underestimated severalfold if we simply adopted the $R_{e, {\rm CO}}$ without any corrections.} Thus, if the spatial extents of the star-forming region and the molecular gas reservoir are largely different, {especially for the galaxies taken from the literature, the comparison presented in Figure \ref{fig:Fig6} (right) might not be straightforward. Indeed, we show in Section \ref{subsec:Subsec3.3} that the effective radii of both the CO line and 870 $\mu$m continuum emissions of the CO emitters are different from the 1.6 $\mu$m radii by factors of {$1.7\pm0.6$ and $4.9\pm2.7$}, respectively.}

%留意事項2:conversion factorの話
Another important caveat is the CO-to-$\rm H_{2}$ conversion factor $\alpha_{\rm CO}$, which depends on the physical conditions of the ISM, especially metallicities. The theoretical models predict that the conversion factor exhibits an upturn in metal-poor environments ($Z\lesssim0.5Z_{\rm\odot}$), where photodissociation plays a significant role in destructing CO molecules \citep{2013ARA&A..51..207B}. \cite{2019A&A...626A..14M} {use H$\alpha$ and [N{\sc ii}] $\lambda$6584 line fluxes to measure the metallicity of the massive galaxies in XCS J2215 whose stellar masses are comparable to the CO emitters.} They find evidence of metallicity enhancement of $\sim$ 0.1 dex compared to coeval star-forming galaxies, which corresponds to $12+\rm{log(O/H)}=8.5$--$8.7$. {By adopting $12+\rm{log(O/H)}=8.5$ and substituting it into Equation (6) of \cite{2015ApJ...800...20G},} which incorporates the effect of the gas-phase metallicity on the conversion factor, i.e. $\alpha(Z)\propto\chi(Z)\alpha_{\rm MW}$, we obtain $\chi(Z)=1.14$. This suggests that the galaxies in XCS J2215 reside in an environment where the metallicity is moderately enhanced, justifying the conversion factor used in this study. Incidentally, most high-redshift star-forming galaxies have a flat metallicity gradient (e.g. \citealp{2016ApJ...827...74W}; \citealp{2020MNRAS.492..821C}; \citealp{2021ApJ...923..203S}). Therefore, the conversion factor is likely uniform across the entire galaxy disk of the CO emitters. For a fair comparison, we use the surface densities of the molecular gas mass derived by using the galactic value of $\alpha_{\rm CO}$ for all of the galaxies shown in Figure \ref{fig:Fig6} (right).

{Nevertheless, it is noteworthy to test the effect on the KS relation, if we adopt a different value for the conversion factor. \cite{2018ApJ...863...56C} present a dynamical constraint on the conversion factor for $z=2-3$ SMGs by using the CO~$J$~=~3--2 line, finding values around $\alpha_{\rm CO}=1.0M_{\rm\odot}$ $\rm(K \ km \ s^{-1} \ pc^{-2})^{-1}$, with an upper limit of $1.4M_{\rm\odot}$ $\rm(K \ km \ s^{-1} \ pc^{-2})^{-1}$. We show the 0.64~dex shift in the right panel of Figure \ref{fig:Fig6} with a leftward arrow, if we adopt a value of $\alpha_{\rm CO}=1.0M_{\rm\odot}$ $\rm(K \ km \ s^{-1} \ pc^{-2})^{-1}$ instead of the galactic value. Both the extended and central regions will then be consistent with local starbursts, and this seems to be plausible given the conditions for starbursts in the high-redshift SMG population.}

\subsection{Structure Formation of Cluster Galaxies at z=1.46}\label{subsec:Subsec4.2}

%P1:Figure7(left)の説明
We now turn our discussion topic to the structure formation of star-forming galaxies in high-redshift clusters. First, we focus on the 1.6 $\rm\mu m$ sizes of the cluster galaxies in XCS J2215. The relation between the stellar mass and the 1.6 $\mu$m size for 17 CO emitters and 14 passive members in XCS J2215 is presented in Figure \ref{fig:Fig7} (left). The 1.6 $\mu$m sizes of these passive members are taken from \cite{2018ApJ...856....8C}. We also show the best-fit mass-size relations of both star-forming and passive galaxies at $z=1.5$, taken from \cite{2014ApJ...788...28V}. We derive these relations by taking the mean of the parameterized fits between $z=1.25$ and $z=1.75$ presented in \cite{2014ApJ...788...28V}. This can be interpreted as the representative mass-size relation at $z=1.5$, since their sample includes both cluster and field galaxies. 

%P2:1.6μmサイズの環境依存性
As seen in Figure \ref{fig:Fig7} (left), the 1.6 $\mu$m sizes of the passive members are systematically smaller than the ones of passive galaxies in field environments. The 1.6 $\mu$m sizes of the majority of CO emitters lie between the correlation of star-forming and passive galaxies at $z=1.5$. Given that 16 of the 17 CO emitters are classified as star-forming members, the distribution of the 1.6 $\mu$m sizes suggests that the gas-rich star-forming cluster galaxies in XCS J2215 are more compact than star-forming field galaxies at $z=1.5$ on average. Nevertheless, care must be taken in drawing a conclusion, since the relations from \cite{2014ApJ...788...28V} are derived from galaxies with a certain redshift range. Therefore, while the optical radii of both star-forming and passive galaxies may or may not depend on the environments in general (\citealp{2019MNRAS.484..595M} and references therein), we tentatively suggest that our CO emitters and passive members in XCS J2215 are more compact than coeval field galaxies.

%P3:Figure(right)の説明
In Figure \ref{fig:Fig7} (right), we compare the effective radii of the HST/1.6 $\mu$m, CO~$J$~=~2--1 line, and 870 $\mu$m continuum of the CO emitters. All three sizes are shown for nine CO emitters, while the 1.6 $\mu$m and CO sizes are shown if they are available for the rest of eight CO emitters. As expected from the size ratio presented in Section \ref{subsec:Subsec3.3}, the three radii show different trends in mass-size distribution. The most extended is the 1.6 $\mu$m size, followed by the CO and 870 $\mu$m sizes.

%P4: 1.6μm vs CO(2-1)
{The average size ratio with standard deviation between the 1.6 $\mu$m and CO sizes of the CO emitters is $1.7\pm0.6$.} The absence of a clear stellar mass dependency indicates that the ratio of the 1.6 $\mu$m and CO sizes is roughly constant as a function of the stellar mass. In general, the 1.6 $\mu$m emission from galaxies traces the stellar distribution. However, because of the radial color gradients derived from the stellar age, metallicity, and dust extinction, the 1.6 $\mu$m size is systematically overestimated against the true size of the stellar component. \cite{2019ApJ...877..103S} measure the half-light radii at 1.6 $\mu$m and half-mass radii for $\sim 7000$ galaxies at $1\leq z\leq2.5$, and find that both the star-forming and passive galaxies have a negative color gradient. Also, they argue that the size ratio of the half-light and half-mass radii evolves with decreasing redshift. For star-forming galaxies with $M_{\star}=10^{10.5} M_{\odot}$ at $z=1.5$, they obtain $\langle r_{\rm light}/r_{\rm mass}\rangle =1.3$. Therefore, we expect that the half-mass radius of the stellar component to be slightly larger than the CO sizes, and it may be characterized by an exponential disk model (Section \ref{subsec:Subsec3.2}).

%P5:870μm vs CO(2-1)
In contrast to the mass-size distributions of the 1.6 $\mu$m and CO~$J$~=~2--1 lines, we find that the 870 $\mu$m sizes of the CO emitters increase with stellar mass (Figure \ref{fig:Fig7}, right). As a consequence, the segregation between the CO~$J$~=~2--1 and 870 $\mu$m sizes becomes less pronounced at the massive end. \cite{2019MNRAS.488.1779C} and \cite{2022MNRAS.510.3321P} conducted FIRE-2 and TNG50 simulations, respectively, to quantify the sizes of the dust in simulated galaxies at different redshifts. Our results are consistent with both simulations, predicting that dust continuum emissions tend to be compact relative to the cold-gas component.

%P5-2:870μm vs CO(2-1) -- inside-out quenching
The compact star formation seen in the nine CO emitters suggests a rapid increase in stellar mass at the center of the extended gas reservoir. Assuming a negligible contribution from radial gas transportation, all nine CO emitters will rapidly exhaust their gas within a typical timescale of $\langle \tau_{\rm cen} \rangle=0.48\pm0.27$ Gyr (Section \ref{subsec:Subsec4.1}), starting from the central region. This leads to a suppression of star formation at the center, which will likely then propagate toward the outer disks within the gas depletion timescale of the extended region. Such a process is in qualitative agreement with the so-called ``inside-out quenching'' scenario (e.g. \citealp{2015Sci...348..314T}; \citealp{2019ApJ...883...81S}).

%P6:870μm (CO emitters) vs 1.6μm (passive) -- transition from star-forming to passive members
The CO emitters will likely become passive cluster members by $z=1.27$ (the redshift after $\sim0.5$ Gyr from $z=1.46$), after depleting most of the molecular gas in the central region. Simultaneously, the 1.6 $\mu$m sizes will be reduced as a consequence of the central star formation activity and the formation of a concentrated stellar component. In turn, the mass-size distribution of the CO emitters will likely become similar to the passive members \citep{2020ApJ...901...74T}. Figure \ref{fig:Fig7} (right) shows that the mass-size distributions of the 870 $\mu$m continuum sizes of the nine CO emitters follow the same trend as the best-fit mass-size relation of the passive members at $1.6$ $\mu$m. This may suggest that the nine CO emitters are experiencing a transition phase, from star-forming to passive members, which in turn may indicate the formation of a bulge-dominated structure, as seen in other $z\sim1$ cluster galaxies (e.g. \citealp{2004MNRAS.350.1005K}; \citealp{2017MNRAS.465L.104N}). Due to the radial color gradient, the half-mass radii of passive galaxies are expected to be smaller than the 1.6 $\mu$m sizes by a similar factor as the star-forming galaxies \citep{2019ApJ...877..103S}.
Moreover, we find that the distributions of the 1.6 $\mu$m sizes of the CO emitters lie between the correlation of the star-forming and passive populations (Figure \ref{fig:Fig7}, left). \cite{2020MNRAS.493.6011M} find a similar mass-size distribution for poststarburst galaxies at $z\sim1$ and argue that the rapid size growth of the passive galaxies can be explained by a combination of minor mergers and newly quenched star formation galaxies. Therefore, the observed 1.6 $\mu$m sizes of the CO emitters further support the scenario in which the CO emitters will rapidly evolve to passive members and populate the massive end of the stellar mass-size relation of passive galaxies.

%P7:cluster scaleのinside-out quenching
Furthermore, in tandem with the 2D distribution map of the CO emitters and passive members (Figure \ref{fig:Fig1}), this is consistent with the picture that quenching propagates from the cluster center to the outskirts. Both passive members will likely evolve into red massive galaxies at $z=0$.

%P8:AGN
Finally, we find that two AGNs, ALMA.11 and ALMA.14, have the second-smallest (1.13 kpc) and smallest (0.98 kpc) CO sizes among the 15 CO emitters. Galaxy-galaxy mergers are thought to generate gas inflows, and feed the supermassive black holes, and eventually trigger AGNs (e.g. \citealp{2005Natur.433..604D}; \citealp{2006ApJS..163....1H}). Moreover, \cite{2010ApJ...722.1666W} point out that mergers can produce compact galaxies with $\sim 1$ kpc sizes. The compactness of ALMA.11 and ALMA.14 may suggest a past episode of a galaxy merger.

\subsection{{Effects of Mergers on Star Formation Activity}}\label{subsec:Subsec4.3}

%P1:イントロ
The enhancement of merger fractions in (proto-)cluster environments has been suggested in various observational studies (e.g. \citealp{2013ApJ...773..154L}; \citealp{2016MNRAS.455.2363H}; \citealp{2018MNRAS.479..703C}; \citealp{2019ApJ...874...63W}). On the contrary, \cite{2017ApJ...843..126D} report no merger enhancement at the core of clusters ($1.51<z<1.74$), although their measurement of the merger fraction in the field region is extraordinarily high (47\%) {compared} to that of other studies, while the merger fraction of the cluster environments is comparable to others. Their high merger fraction in the $z\sim1.6$ field is likely due to the small sample size of the control sample. Our visual inspection of early-stage mergers toward the 17 CO emitters (Section \ref{subsec:Subsec3.1}) reveals {an upper limit on the merger fraction of 35\% (6/17)}, which is a factor of 3 higher than the value of coeval field galaxies (e.g. $\sim$11\% at $z=1.62$ in \citealp{2013ApJ...773..154L}). Therefore, although the parent sample of this calculation is biased toward gas-rich systems, it is worth investigating whether these early-stage mergers (with a projected separation within 15 kpc) play a role in their star formation properties and {structures}.

%P2:no difference between merger/nonmerger
We inspect whether the effective radii (Figure \ref{fig:Fig5}) and the KS relation (Figure \ref{fig:Fig6} left) are different between early-stage mergers and nonmergers. In both figures, we do not find any systematic differences between the two populations, implying that early-stage mergers are not an important factor in determining {the spatial extent of the CO~$J$~=~2--1 line and 870 $\rm \mu m$ continuum emissions, nor the star formation regime}. Both the simulation and observation results in \cite{2012MNRAS.426..549S} present an enhanced SFR by $\sim$0.21 dex within a separation of $\sim$30 kpc during an early-stage major merger. However, this study uses Sloan Digital Sky Survey galaxies with $z<0.16$ as their sample, thus their results may not be applicable to high-redshift mergers. It is known that the gas mass fraction increases as a function of redshift \citep{2020ARA&A..58..157T}, motivating collision simulations of gas-rich disk galaxies as a proxy of high-redshift mergers (\citealp{2014A&A...562A...1P}; \citealp{2015MNRAS.449.3719S}; \citealp{2017MNRAS.465.1934F}). All these simulations show that high-redshift gas-rich mergers do not trigger SFR enhancement as much as low-redshift mergers. 
Observations of early-stage mergers at $z>1$ also agree with the absence of SFR enhancement (\citealp{2013MNRAS.429L..40K}; \citealp{2018ApJ...868...46S}; \citealp{2019MNRAS.485.5631C}). 
The surface densities of the molecular gas mass and SFR derived in this study provide further supporting evidence that there is no SFR enhancement at the central region for early-stage mergers when compared to nonmergers. Therefore, from our results, we confirm the absence of merger dependency on star formation properties, even in the resolved view.

%ALMA.05: hidden starburst?
The SFR enhancement due to a merger event is expected to reach a peak at its late stage (e.g. \citealp{2012MNRAS.426..549S}). However, it is difficult to identify late-stage mergers even if the emissions are resolved, because the projection effect makes it increasingly difficult to disentangle the two galaxies involved. In Figure \ref{fig:Fig6}, one galaxy, ALMA.05{,} has an extremely short gas depletion timescale in the central region ($\sim 70$ Myr), while that in the extended region ($\sim 1.4$ Gyr) is comparable to the others. This is due to the compactness of the 870 $\rm\mu m$ emission, and suggests that ALMA.05 is hosting a starburst in the central region, which is more extreme than the others. As all of the nine CO emitters lie in the star-forming MS at $z=1.46$ (Figure \ref{fig:Fig2}), ALMA.05 is likely to be a starburst ``hidden'' in the star-forming MS \citep{2018A&A...616A.110E}. Following \cite{2021MNRAS.508.5217P}, {we therefore suggest that ALMA.05 is a poststarburst galaxy, presumably driven by a merger event in the past. To further confirm this scenario, it is necessary to examine the CO excitation by taking the luminosity ratio of the CO~$J$~=~2--1 and mid-/high-$J$ transition lines, under the assumption that poststarburst phases that have experienced mergers have higher excitation \citep{2021MNRAS.508.5217P}.}

Recently, the ALMaQUEST survey \citep{2020ApJ...903..145L} has revealed that the resolved MS relation in nearby galaxies can be explained as {a consequence of the combination} of the resolved KS relation and the resolved $M_{\star}$-$M_{\rm mol \ gas}$ relation (\citealp{2019ApJ...884L..33L}; \citealp{2021MNRAS.501.4777E}). ALMA.05 provides a clear example of how the resolved KS relation can be used to detect such starbursts hidden in the star-forming MS. 
\section{Conclusions} \label{sec:Conclusions}
We have studied 17 star-forming galaxies associated with XCS J2215 at $z=1.46$ by analysing the $0\farcs4$ resolution CO~$J$~=~2--1 line emission and the $0\farcs2$ resolution 870 $\mu$m continuum emission. The main goal of this study was to measure the effective radii by fitting the visibility data, and we have successfully measured the effective radii of the dust-obscured star formation for nine galaxies and of the cold {molecular gas for} 15 galaxies. Our major findings are summarized below.

\begin{enumerate}
\item{\textit{The compactness of the dust emission.} All of the nine size-measured galaxies show relatively compact dust emission compared to the CO~$J$~=~2--1 line emission by a factor of {$2.8\pm1.4$} (Figure \ref{fig:Fig5}). The average sizes are $R_{e,{\rm CO}}=1.82\pm0.48$ kpc and $R_{e,870{\rm\mu m}}=0.79\pm0.47$} kpc.
\item{\textit{Enhanced star formation in the central region.} We demonstrate the resolved KS relation for the nine CO emitters by dividing each galaxy into two subregions. We find a trend that the central region has a shorter gas depletion timescale of $\langle \tau_{\rm cen} \rangle=0.48\pm0.27$ Gyr compared to the extended region, which is consistent with the compactness of the dust emission (Figure \ref{fig:Fig6}). This can be {attributed} to the enhanced star formation activity in the central {$\sim1$ kpc region} of the galaxies.}
\item{\textit{Evidence of transition from star-forming to passive members.} We find an agreement between the 870 $\mu$m radii of the CO emitters and the 1.6 $\mu$m radii of the passive members in the stellar mass-size distribution (Figure \ref{fig:Fig7}, {right}). This suggests that the star-forming regions in the CO emitters will become concentrated stellar bulge components, which are expected to be seen in passive galaxies by $z=1.27$ ($\sim 0.5$ Gyr after $z=1.46$)}.
\item{\textit{No effect of early-stage mergers.}; From the {CLEANed images} of the CO~$J$~=~2--1 line emission, we classify 17 CO emitters into early-stage mergers and nonmergers, with six of them being found to be {early-stage} mergers. We do not find a significant difference {in the sizes measured by ALMA and in the spatially resolved star formation properties} between these six mergers and other galaxies. This implies that early-stage mergers {do not play an important role in} high-density environments, as predicted by gas-rich merger simulations.}
\end{enumerate}

%Future prospects
To examine whether our findings are unique to cluster galaxies at high redshift, it is necessary to expand similar {high-resolution} observations toward coeval field galaxies. Moreover, {it will be possible to} expand our analysis toward (proto-)clusters at higher redshift ($z\gtrsim2$) when ALMA Band 1 (35--50 GHz) and Band 2 (65--90 GHz) become available. These new receiver bands will enable us to probe in high resolution parts of the more distant universe that have never been reached before. 

\begin{acknowledgments}
We thank the anonymous referee and editor for their comments that improved the article. This paper makes use of the following ALMA data: ADS/JAO.ALMA \#2012.1.00623.1, \#2015.1.00779.S, \#2017.1.00471.S, and \#2017.1.01228.S. ALMA is a partnership of ESO (representing its member states), NSF (USA), and NINS (Japan), together with NRC (Canada), MOST and ASIAA (Taiwan), and KASI (Republic of Korea), in cooperation with the Republic of Chile. The Joint ALMA Observatory is operated by ESO, AUI/NRAO, and NAOJ. We thank the ALMA staff and, in particular, the EA-ARC staff for their support. 
K.T. acknowledges the support from the JSPS Grant-in-Aid for Scientific Research (C) 20K14526. 
D.I. is supported by JSPS KAKENHI grant No. JP21H01133. T.K. acknowledges the support from the Grant-in-Aid for Scientific Research (A) (KAKENHI \#18H03717) and from the NAOJ ALMA Scientific Research grant No. 2018-08A. 
T.I. acknowledges the support from the JSPS KAKENHI grant No. JP20K14531. 
K.K. acknowledges the support from the JSPS KAKENHI grant No. JP17H06130 and from the NAOJ ALMA Scientific Research grant No. 2017-06B. 
Data analyses were carried out in part on the Multi-wavelength Data Analysis System operated by the Astronomy Data Center (ADC) at the National Astronomical Observatory of Japan. \\
\end{acknowledgments}

\begin{appendix}
\restartappendixnumbering
\section{The spectra of the CO~$J$~=~2--1  line}\label{sec:CO (2-1)spectra}

\begin{figure*}[h]
\epsscale{1.1}
\plotone{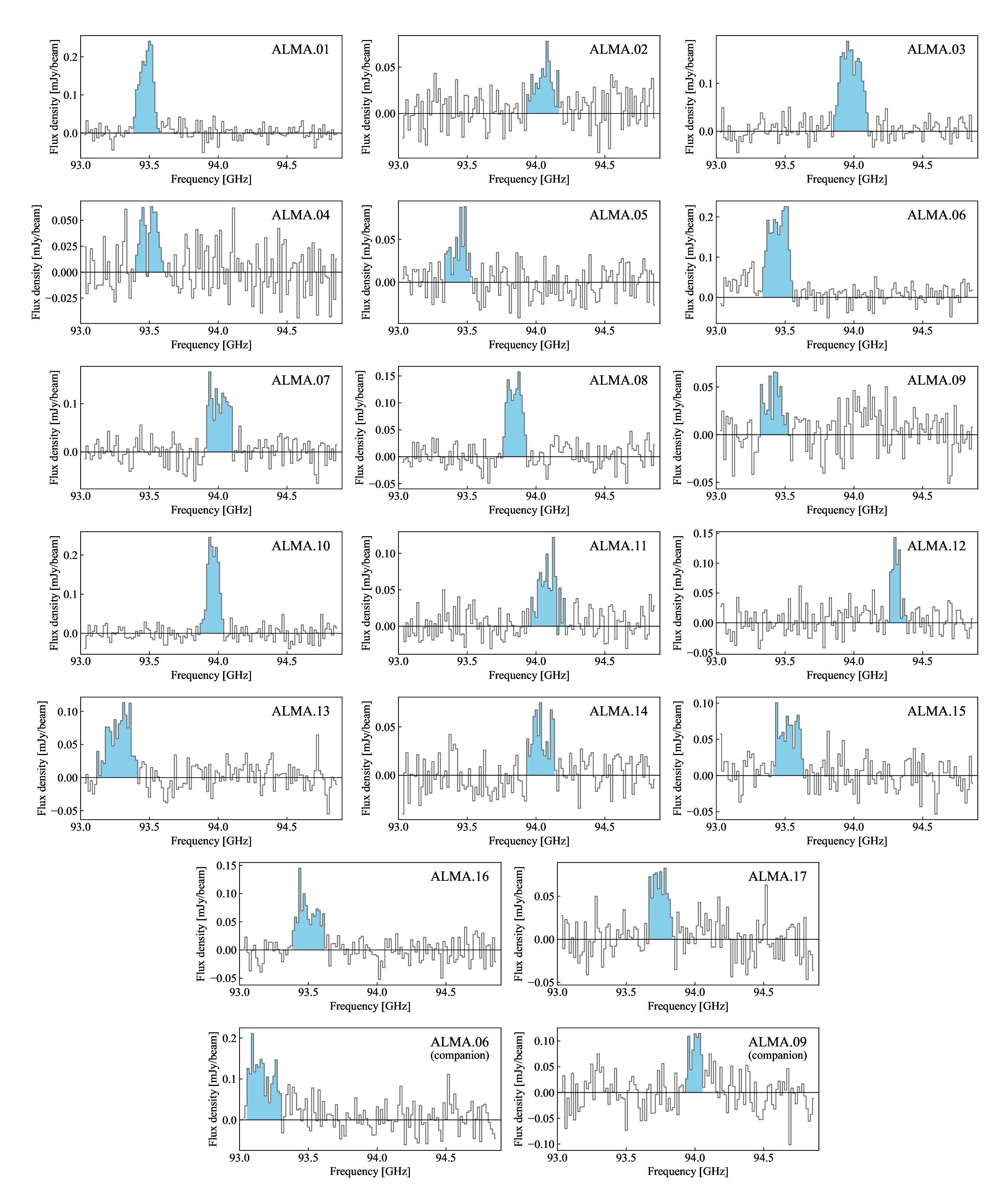}
\caption{The CO~$J$~=~2--1 spectra of 17 CO emitters and two companion galaxies (bottom row). The spectra are taken from the clean images created by concatenated data from Cycle 3 and Cycle 5. The spectral resolution is 50~km~s$^{-1}$. We take a $2\farcs0$ aperture for the main components of the CO emitters and a $1\farcs0$ aperture for the companions. The filled bins in light blue are the frequency ranges that we select for both imaging and visibility analyses. \label{fig:Fig8}}
\end{figure*}

The spectra covering the CO~$J$~=~2--1 line in 17 CO emitters and two companion galaxies are shown in Figure \ref{fig:Fig8}.

\section{Visibility amplitudes}\label{sec:uvamp}
\restartappendixnumbering

\begin{figure*}[ht!]
\epsscale{1.1}
\plotone{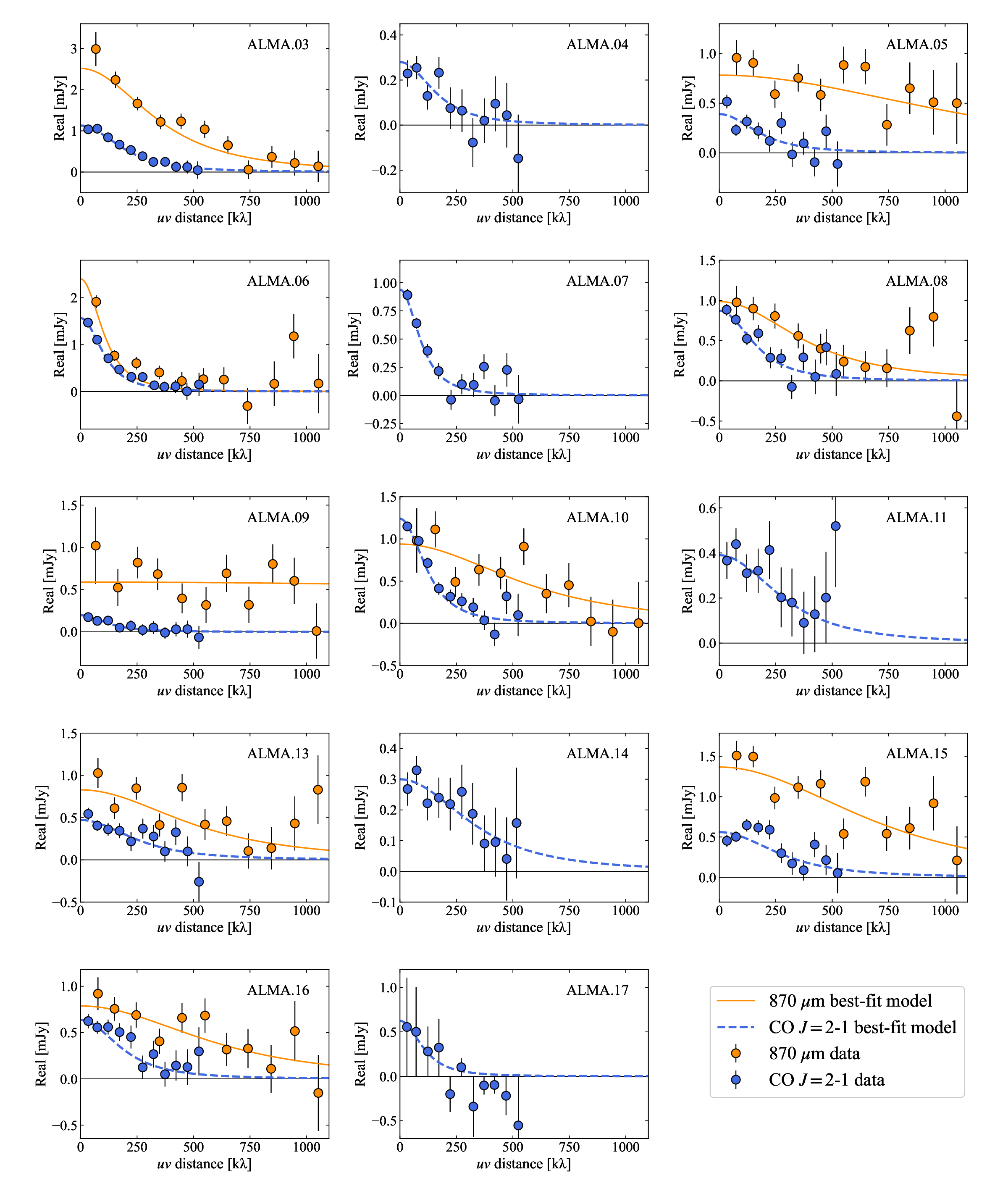}
\caption{The amplitudes of the visibility data as a function of {\textit uv} distance. The averaged amplitudes in each bin are shown as filled circles (blue: CO~$J$~=~2--1 line; orange: 870 $\mu$m continuum). We take 50 and 100 k$\lambda$ bins in the CO~$J$~=~2--1 line and 870 $\mu$m continuum  emissions, respectively. The descending curves along the {\textit uv} distance signify the best-fit exponential disk model of each emission. \label{fig:Fig9}}
\end{figure*}

We show the real part of visibility data as a function of {\textit uv} distance for 14 CO emitters in Figure \ref{fig:Fig9}.

\section{Comparison of CO emitters in different clusters}\label{sec:HST_CO}
\begin{figure}[!h]
\epsscale{1.2}
\plotone{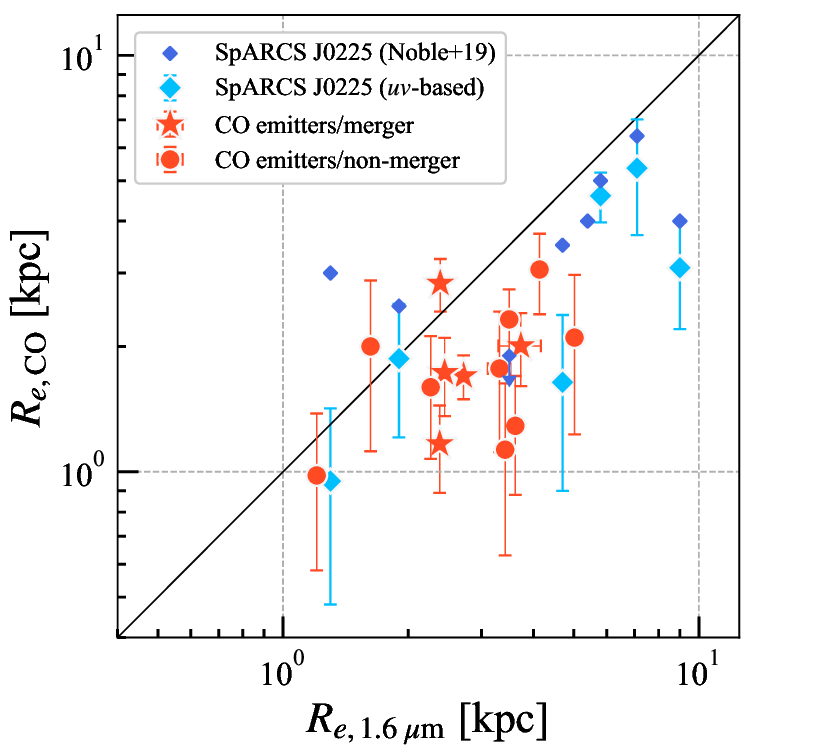}
\caption{{Comparison of the 1.6 $\mu$m and CO radii. The black solid line corresponds to the equality of both radii. The orange stars and circles denote the mergers and nonmergers in XCS J2215, respectively. ALMA.02, ALMA.09, and ALMA.12 are excluded from this figure, since neither 1.6 $\mu$m nor CO sizes are available. The blue diamonds with an error bar correspond to the CO emitters in SpARCS J0225, in which the CO sizes are measured based on {\textit uv} fittings. The CO sizes derived from the CLEANed images (the blue diamonds without an error bar) are also shown for reference. For the CO emitters in SpARCS J0225, we take the 1.6 $\mu$m sizes from the values tabulated in \cite{2019ApJ...870...56N}.}} \label{fig:Fig10}
\end{figure}

\restartappendixnumbering
{In order to compare with the CO emitters in XCS J2215 fairly, we measure the effective radii of the CO~$J$~=~2--1 line emissions of eight SpARCS J0225 galaxies reported in \cite{2019ApJ...870...56N}, in the same manner as described in Section \ref{subsec:Subsec3.2}. As a result, we successfully measure six CO sizes. Figure \ref{fig:Fig10} illustrates the comparison of the 1.6 $\mu$m and CO sizes of the CO emitters in XCS J2215 or SpARCS J0225. For both galaxy populations, the 1.6 $\mu$m sizes tend to be larger than the CO sizes. This trend is more pronounced in the {\textit uv}-based measurements than the image-based measurements for SpARCS J0225, since the CO sizes are systematically smaller in the {\textit uv}-based measurements. Compared to the nine CO emitters in XCS J2215, three of the six CO emitters in SpARCS J0225 are spatially larger in both radii.}

\end{appendix}

\bibliographystyle{apj.bst}
\bibliography{Ikeda_2215.bib}

\begin{thebibliography}{}
\expandafter\ifx\csname natexlab\endcsname\relax\def\natexlab#1{#1}\fi

\bibitem[{{Alberts} {et~al.}(2022){Alberts}, {Adams}, {Gregg}, {Pope},
  {Williams}, \& {Eisenhardt}}]{2022ApJ...927..235A}
{Alberts}, S., {Adams}, J., {Gregg}, B., {et~al.} 2022, \apj, 927, 235

\bibitem[{{Alberts} {et~al.}(2014){Alberts}, {Pope}, {Brodwin}, {Atlee}, {Lin},
  {Dey}, {Eisenhardt}, {Gettings}, {Gonzalez}, {Jannuzi}, {Mancone},
  {Moustakas}, {Snyder}, {Stanford}, {Stern}, {Weiner}, \&
  {Zeimann}}]{2014MNRAS.437..437A}
{Alberts}, S., {Pope}, A., {Brodwin}, M., {et~al.} 2014, \mnras, 437, 437

\bibitem[{{Aoyama} {et~al.}(2022){Aoyama}, {Kodama}, {Suzuki}, {Tadaki},
  {Shimakawa}, {Hayashi}, {Koyama}, \&
  {P{\'e}rez-Mart{\'\i}nez}}]{2022ApJ...924...74A}
{Aoyama}, K., {Kodama}, T., {Suzuki}, T.~L., {et~al.} 2022, \apj, 924, 74

\bibitem[{{Bamford} {et~al.}(2009){Bamford}, {Nichol}, {Baldry}, {Land},
  {Lintott}, {Schawinski}, {Slosar}, {Szalay}, {Thomas}, {Torki}, {Andreescu},
  {Edmondson}, {Miller}, {Murray}, {Raddick}, \&
  {Vandenberg}}]{2009MNRAS.393.1324B}
{Bamford}, S.~P., {Nichol}, R.~C., {Baldry}, I.~K., {et~al.} 2009, \mnras, 393,
  1324

\bibitem[{{Beifiori} {et~al.}(2017){Beifiori}, {Mendel}, {Chan}, {Saglia},
  {Bender}, {Cappellari}, {Davies}, {Galametz}, {Houghton}, {Prichard},
  {Smith}, {Stott}, {Wilman}, {Lewis}, {Sharples}, \&
  {Wegner}}]{2017ApJ...846..120B}
{Beifiori}, A., {Mendel}, J.~T., {Chan}, J. C.~C., {et~al.} 2017, \apj, 846,
  120

\bibitem[{{Bigiel} {et~al.}(2008){Bigiel}, {Leroy}, {Walter}, {Brinks}, {de
  Blok}, {Madore}, \& {Thornley}}]{2008AJ....136.2846B}
{Bigiel}, F., {Leroy}, A., {Walter}, F., {et~al.} 2008, \aj, 136, 2846

\bibitem[{{Bigiel} {et~al.}(2011){Bigiel}, {Leroy}, {Walter}, {Brinks}, {de
  Blok}, {Kramer}, {Rix}, {Schruba}, {Schuster}, {Usero}, \&
  {Wiesemeyer}}]{2011ApJ...730L..13B}
{Bigiel}, F., {Leroy}, A.~K., {Walter}, F., {et~al.} 2011, \apjl, 730, L13

\bibitem[{{Blanton} {et~al.}(2005){Blanton}, {Eisenstein}, {Hogg}, {Schlegel},
  \& {Brinkmann}}]{2005ApJ...629..143B}
{Blanton}, M.~R., {Eisenstein}, D., {Hogg}, D.~W., {Schlegel}, D.~J., \&
  {Brinkmann}, J. 2005, \apj, 629, 143

\bibitem[{{Bolatto} {et~al.}(2013){Bolatto}, {Wolfire}, \&
  {Leroy}}]{2013ARA&A..51..207B}
{Bolatto}, A.~D., {Wolfire}, M., \& {Leroy}, A.~K. 2013, \araa, 51, 207

\bibitem[{{Boogaard} {et~al.}(2019){Boogaard}, {Decarli},
  {Gonz{\'a}lez-L{\'o}pez}, {van der Werf}, {Walter}, {Bouwens}, {Aravena},
  {Carilli}, {Bauer}, {Brinchmann}, {Contini}, {Cox}, {da Cunha}, {Daddi},
  {D{\'\i}az-Santos}, {Hodge}, {Inami}, {Ivison}, {Maseda}, {Matthee}, {Oesch},
  {Popping}, {Riechers}, {Schaye}, {Schouws}, {Smail}, {Weiss}, {Wisotzki},
  {Bacon}, {Cortes}, {Rix}, {Somerville}, {Swinbank}, \&
  {Wagg}}]{2019ApJ...882..140B}
{Boogaard}, L.~A., {Decarli}, R., {Gonz{\'a}lez-L{\'o}pez}, J., {et~al.} 2019,
  \apj, 882, 140

\bibitem[{{Brodwin} {et~al.}(2013){Brodwin}, {Stanford}, {Gonzalez}, {Zeimann},
  {Snyder}, {Mancone}, {Pope}, {Eisenhardt}, {Stern}, {Alberts}, {Ashby},
  {Brown}, {Chary}, {Dey}, {Galametz}, {Gettings}, {Jannuzi}, {Miller},
  {Moustakas}, \& {Moustakas}}]{2013ApJ...779..138B}
{Brodwin}, M., {Stanford}, S.~A., {Gonzalez}, A.~H., {et~al.} 2013, \apj, 779,
  138

\bibitem[{{Calistro Rivera} {et~al.}(2018){Calistro Rivera}, {Hodge}, {Smail},
  {Swinbank}, {Weiss}, {Wardlow}, {Walter}, {Rybak}, {Chen}, {Brandt},
  {Coppin}, {da Cunha}, {Dannerbauer}, {Greve}, {Karim}, {Knudsen},
  {Schinnerer}, {Simpson}, {Venemans}, \& {van der Werf}}]{2018ApJ...863...56C}
{Calistro Rivera}, G., {Hodge}, J.~A., {Smail}, I., {et~al.} 2018, \apj, 863,
  56

\bibitem[{{Chabrier}(2003)}]{2003PASP..115..763C}
{Chabrier}, G. 2003, \pasp, 115, 763

\bibitem[{{Champagne} {et~al.}(2021){Champagne}, {Casey}, {Zavala}, {Cooray},
  {Dannerbauer}, {Fabian}, {Hayward}, {Long}, \&
  {Spilker}}]{2021ApJ...913..110C}
{Champagne}, J.~B., {Casey}, C.~M., {Zavala}, J.~A., {et~al.} 2021, \apj, 913,
  110

\bibitem[{{Chan} {et~al.}(2016){Chan}, {Beifiori}, {Mendel}, {Saglia},
  {Bender}, {Fossati}, {Galametz}, {Wegner}, {Wilman}, {Cappellari}, {Davies},
  {Houghton}, {Prichard}, {Lewis}, {Sharples}, \&
  {Stott}}]{2016MNRAS.458.3181C}
{Chan}, J. C.~C., {Beifiori}, A., {Mendel}, J.~T., {et~al.} 2016, \mnras, 458,
  3181

\bibitem[{{Chan} {et~al.}(2018){Chan}, {Beifiori}, {Saglia}, {Mendel}, {Stott},
  {Bender}, {Galametz}, {Wilman}, {Cappellari}, {Davies}, {Houghton},
  {Prichard}, {Lewis}, {Sharples}, \& {Wegner}}]{2018ApJ...856....8C}
{Chan}, J. C.~C., {Beifiori}, A., {Saglia}, R.~P., {et~al.} 2018, \apj, 856, 8

\bibitem[{{Chen} {et~al.}(2017){Chen}, {Hodge}, {Smail}, {Swinbank}, {Walter},
  {Simpson}, {Calistro Rivera}, {Bertoldi}, {Brandt}, {Chapman}, {da Cunha},
  {Dannerbauer}, {De Breuck}, {Harrison}, {Ivison}, {Karim}, {Knudsen},
  {Wardlow}, {Wei{\ss}}, \& {van der Werf}}]{2017ApJ...846..108C}
{Chen}, C.-C., {Hodge}, J.~A., {Smail}, I., {et~al.} 2017, \apj, 846, 108

\bibitem[{{Chiang} {et~al.}(2013){Chiang}, {Overzier}, \&
  {Gebhardt}}]{2013ApJ...779..127C}
{Chiang}, Y.-K., {Overzier}, R., \& {Gebhardt}, K. 2013, \apj, 779, 127

\bibitem[{{Chiang} {et~al.}(2017){Chiang}, {Overzier}, {Gebhardt}, \&
  {Henriques}}]{2017ApJ...844L..23C}
{Chiang}, Y.-K., {Overzier}, R.~A., {Gebhardt}, K., \& {Henriques}, B. 2017,
  \apjl, 844, L23

\bibitem[{{Cibinel} {et~al.}(2019){Cibinel}, {Daddi}, {Sargent}, {Le Floc'h},
  {Liu}, {Bournaud}, {Oesch}, {Amram}, {Calabr{\`o}}, {Duc}, {Pannella},
  {Puglisi}, {Perret}, {Elbaz}, \& {Kokorev}}]{2019MNRAS.485.5631C}
{Cibinel}, A., {Daddi}, E., {Sargent}, M.~T., {et~al.} 2019, \mnras, 485, 5631

\bibitem[{{Cochrane} {et~al.}(2019){Cochrane}, {Hayward},
  {Angl{\'e}s-Alc{\'a}zar}, {Lotz}, {Parsotan}, {Ma}, {Kere{\v{s}}},
  {Feldmann}, {Faucher-Gigu{\`e}re}, \& {Hopkins}}]{2019MNRAS.488.1779C}
{Cochrane}, R.~K., {Hayward}, C.~C., {Angl{\'e}s-Alc{\'a}zar}, D., {et~al.}
  2019, \mnras, 488, 1779

\bibitem[{{Coogan} {et~al.}(2018){Coogan}, {Daddi}, {Sargent}, {Strazzullo},
  {Valentino}, {Gobat}, {Magdis}, {Bethermin}, {Pannella}, {Onodera}, {Liu},
  {Cimatti}, {Dannerbauer}, {Carollo}, {Renzini}, \&
  {Tremou}}]{2018MNRAS.479..703C}
{Coogan}, R.~T., {Daddi}, E., {Sargent}, M.~T., {et~al.} 2018, \mnras, 479, 703

\bibitem[{{Cooke} {et~al.}(2019){Cooke}, {Smail}, {Stach}, {Swinbank}, {Bower},
  {Chen}, {Koyama}, \& {Thomson}}]{2019MNRAS.486.3047C}
{Cooke}, E.~A., {Smail}, I., {Stach}, S.~M., {et~al.} 2019, \mnras, 486, 3047

\bibitem[{{Curti} {et~al.}(2020){Curti}, {Maiolino}, {Cirasuolo}, {Mannucci},
  {Williams}, {Auger}, {Mercurio}, {Hayden-Pawson}, {Cresci}, {Marconi},
  {Belfiore}, {Cappellari}, {Cicone}, {Cullen}, {Meneghetti}, {Ota}, {Peng},
  {Pettini}, {Swinbank}, \& {Troncoso}}]{2020MNRAS.492..821C}
{Curti}, M., {Maiolino}, R., {Cirasuolo}, M., {et~al.} 2020, \mnras, 492, 821

\bibitem[{{da Cunha} {et~al.}(2015){da Cunha}, {Walter}, {Smail}, {Swinbank},
  {Simpson}, {Decarli}, {Hodge}, {Weiss}, {van der Werf}, {Bertoldi},
  {Chapman}, {Cox}, {Danielson}, {Dannerbauer}, {Greve}, {Ivison}, {Karim}, \&
  {Thomson}}]{2015ApJ...806..110D}
{da Cunha}, E., {Walter}, F., {Smail}, I.~R., {et~al.} 2015, \apj, 806, 110

\bibitem[{{Daddi} {et~al.}(2010){Daddi}, {Elbaz}, {Walter}, {Bournaud},
  {Salmi}, {Carilli}, {Dannerbauer}, {Dickinson}, {Monaco}, \&
  {Riechers}}]{2010ApJ...714L.118D}
{Daddi}, E., {Elbaz}, D., {Walter}, F., {et~al.} 2010, \apjl, 714, L118

\bibitem[{{D'Amato} {et~al.}(2020){D'Amato}, {Gilli}, {Prandoni}, {Vignali},
  {Massardi}, {Mignoli}, {Cucciati}, {Morishita}, {Decarli}, {Brusa}, {Calura},
  {Balmaverde}, {Chiaberge}, {Liuzzo}, {Nanni}, {Peca}, {Pensabene}, {Tozzi},
  \& {Norman}}]{2020A&A...641L...6D}
{D'Amato}, Q., {Gilli}, R., {Prandoni}, I., {et~al.} 2020, \aap, 641, L6

\bibitem[{{Dame} {et~al.}(2001){Dame}, {Hartmann}, \&
  {Thaddeus}}]{2001ApJ...547..792D}
{Dame}, T.~M., {Hartmann}, D., \& {Thaddeus}, P. 2001, \apj, 547, 792

\bibitem[{{Dannerbauer} {et~al.}(2017){Dannerbauer}, {Lehnert}, {Emonts},
  {Ziegler}, {Altieri}, {De Breuck}, {Hatch}, {Kodama}, {Koyama}, {Kurk},
  {Matiz}, {Miley}, {Narayanan}, {Norris}, {Overzier}, {R{\"o}ttgering},
  {Sargent}, {Seymour}, {Tanaka}, {Valtchanov}, \&
  {Wylezalek}}]{2017A&A...608A..48D}
{Dannerbauer}, H., {Lehnert}, M.~D., {Emonts}, B., {et~al.} 2017, \aap, 608,
  A48

\bibitem[{{de los Reyes} \& {Kennicutt}(2019)}]{2019ApJ...872...16D}
{de los Reyes}, M. A.~C., \& {Kennicutt}, Robert~C., J. 2019, \apj, 872, 16

\bibitem[{{De Lucia} \& {Blaizot}(2007)}]{2007MNRAS.375....2D}
{De Lucia}, G., \& {Blaizot}, J. 2007, \mnras, 375, 2

\bibitem[{{Decarli} {et~al.}(2019){Decarli}, {Walter},
  {G{\'o}nzalez-L{\'o}pez}, {Aravena}, {Boogaard}, {Carilli}, {Cox}, {Daddi},
  {Popping}, {Riechers}, {Uzgil}, {Weiss}, {Assef}, {Bacon}, {Bauer},
  {Bertoldi}, {Bouwens}, {Contini}, {Cortes}, {da Cunha}, {D{\'\i}az-Santos},
  {Elbaz}, {Inami}, {Hodge}, {Ivison}, {Le F{\`e}vre}, {Magnelli}, {Novak},
  {Oesch}, {Rix}, {Sargent}, {Smail}, {Swinbank}, {Somerville}, {van der Werf},
  {Wagg}, \& {Wisotzki}}]{2019ApJ...882..138D}
{Decarli}, R., {Walter}, F., {G{\'o}nzalez-L{\'o}pez}, J., {et~al.} 2019, \apj,
  882, 138

\bibitem[{{Delahaye} {et~al.}(2017){Delahaye}, {Webb}, {Nantais}, {DeGroot},
  {Wilson}, {Muzzin}, {Yee}, {Foltz}, {Noble}, {Demarco}, {Tudorica}, {Cooper},
  {Lidman}, {Perlmutter}, {Hayden}, {Boone}, \& {Surace}}]{2017ApJ...843..126D}
{Delahaye}, A.~G., {Webb}, T.~M.~A., {Nantais}, J., {et~al.} 2017, \apj, 843,
  126

\bibitem[{{Di Matteo} {et~al.}(2005){Di Matteo}, {Springel}, \&
  {Hernquist}}]{2005Natur.433..604D}
{Di Matteo}, T., {Springel}, V., \& {Hernquist}, L. 2005, \nat, 433, 604

\bibitem[{{Dressler}(1980)}]{1980ApJ...236..351D}
{Dressler}, A. 1980, \apj, 236, 351

\bibitem[{{Dunlop} {et~al.}(2017){Dunlop}, {McLure}, {Biggs}, {Geach},
  {Micha{\l}owski}, {Ivison}, {Rujopakarn}, {van Kampen}, {Kirkpatrick},
  {Pope}, {Scott}, {Swinbank}, {Targett}, {Aretxaga}, {Austermann}, {Best},
  {Bruce}, {Chapin}, {Charlot}, {Cirasuolo}, {Coppin}, {Ellis}, {Finkelstein},
  {Hayward}, {Hughes}, {Ibar}, {Jagannathan}, {Khochfar}, {Koprowski},
  {Narayanan}, {Nyland}, {Papovich}, {Peacock}, {Rieke}, {Robertson},
  {Vernstrom}, {Werf}, {Wilson}, \& {Yun}}]{2017MNRAS.466..861D}
{Dunlop}, J.~S., {McLure}, R.~J., {Biggs}, A.~D., {et~al.} 2017, \mnras, 466,
  861

\bibitem[{{Elbaz} {et~al.}(2018){Elbaz}, {Leiton}, {Nagar}, {Okumura},
  {Franco}, {Schreiber}, {Pannella}, {Wang}, {Dickinson}, {D{\'\i}az-Santos},
  {Ciesla}, {Daddi}, {Bournaud}, {Magdis}, {Zhou}, \&
  {Rujopakarn}}]{2018A&A...616A.110E}
{Elbaz}, D., {Leiton}, R., {Nagar}, N., {et~al.} 2018, \aap, 616, A110

\bibitem[{{Ellison} {et~al.}(2021){Ellison}, {Lin}, {Thorp}, {Pan}, {Scudder},
  {S{\'a}nchez}, {Bluck}, \& {Maiolino}}]{2021MNRAS.501.4777E}
{Ellison}, S.~L., {Lin}, L., {Thorp}, M.~D., {et~al.} 2021, \mnras, 501, 4777

\bibitem[{{Engel} {et~al.}(2010){Engel}, {Tacconi}, {Davies}, {Neri}, {Smail},
  {Chapman}, {Genzel}, {Cox}, {Greve}, {Ivison}, {Blain}, {Bertoldi}, \&
  {Omont}}]{2010ApJ...724..233E}
{Engel}, H., {Tacconi}, L.~J., {Davies}, R.~I., {et~al.} 2010, \apj, 724, 233

\bibitem[{{Fensch} {et~al.}(2017){Fensch}, {Renaud}, {Bournaud}, {Duc},
  {Agertz}, {Amram}, {Combes}, {Di Matteo}, {Elmegreen}, {Emsellem}, {Jog},
  {Perret}, {Struck}, \& {Teyssier}}]{2017MNRAS.465.1934F}
{Fensch}, J., {Renaud}, F., {Bournaud}, F., {et~al.} 2017, \mnras, 465, 1934

\bibitem[{{Fujimoto} {et~al.}(2018){Fujimoto}, {Ouchi}, {Kohno}, {Yamaguchi},
  {Hatsukade}, {Ueda}, {Shibuya}, {Inoue}, {Oogi}, {Toft},
  {G{\'o}mez-Guijarro}, {Wang}, {Espada}, {Nagao}, {Tanaka}, {Ao}, {Umehata},
  {Taniguchi}, {Nakanishi}, {Rujopakarn}, {Ivison}, {Wang}, {Lee}, {Tadaki},
  {Tamura}, \& {Dunlop}}]{2018ApJ...861....7F}
{Fujimoto}, S., {Ouchi}, M., {Kohno}, K., {et~al.} 2018, \apj, 861, 7

\bibitem[{{Fujimoto} {et~al.}(2020){Fujimoto}, {Silverman}, {Bethermin},
  {Ginolfi}, {Jones}, {Le F{\`e}vre}, {Dessauges-Zavadsky}, {Rujopakarn},
  {Faisst}, {Fudamoto}, {Cassata}, {Morselli}, {Maiolino}, {Schaerer}, {Capak},
  {Yan}, {Vallini}, {Toft}, {Loiacono}, {Zamorani}, {Talia}, {Narayanan},
  {Hathi}, {Lemaux}, {Boquien}, {Amorin}, {Ibar}, {Koekemoer},
  {M{\'e}ndez-Hern{\'a}ndez}, {Bardelli}, {Vergani}, {Zucca}, {Romano}, \&
  {Cimatti}}]{2020ApJ...900....1F}
{Fujimoto}, S., {Silverman}, J.~D., {Bethermin}, M., {et~al.} 2020, \apj, 900,
  1

\bibitem[{{Gaia Collaboration} {et~al.}(2021){Gaia Collaboration}, {Brown},
  {Vallenari}, {Prusti}, {de Bruijne}, {Babusiaux}, {Biermann}, {Creevey},
  {Evans}, {Eyer}, {Hutton}, {Jansen}, {Jordi}, {Klioner}, {Lammers},
  {Lindegren}, {Luri}, {Mignard}, {Panem}, {Pourbaix}, {Randich}, {Sartoretti},
  {Soubiran}, {Walton}, {Arenou}, {Bailer-Jones}, {Bastian}, {Cropper},
  {Drimmel}, {Katz}, {Lattanzi}, {van Leeuwen}, {Bakker}, {Cacciari},
  {Casta{\~n}eda}, {De Angeli}, {Ducourant}, {Fabricius}, {Fouesneau},
  {Fr{\'e}mat}, {Guerra}, {Guerrier}, {Guiraud}, {Jean-Antoine Piccolo},
  {Masana}, {Messineo}, {Mowlavi}, {Nicolas}, {Nienartowicz}, {Pailler},
  {Panuzzo}, {Riclet}, {Roux}, {Seabroke}, {Sordo}, {Tanga}, {Th{\'e}venin},
  {Gracia-Abril}, {Portell}, {Teyssier}, {Altmann}, {Andrae}, {Bellas-Velidis},
  {Benson}, {Berthier}, {Blomme}, {Brugaletta}, {Burgess}, {Busso}, {Carry},
  {Cellino}, {Cheek}, {Clementini}, {Damerdji}, {Davidson}, {Delchambre},
  {Dell'Oro}, {Fern{\'a}ndez-Hern{\'a}ndez}, {Galluccio}, {Garc{\'\i}a-Lario},
  {Garcia-Reinaldos}, {Gonz{\'a}lez-N{\'u}{\~n}ez}, {Gosset}, {Haigron},
  {Halbwachs}, {Hambly}, {Harrison}, {Hatzidimitriou}, {Heiter},
  {Hern{\'a}ndez}, {Hestroffer}, {Hodgkin}, {Holl}, {Jan{\ss}en}, {Jevardat de
  Fombelle}, {Jordan}, {Krone-Martins}, {Lanzafame}, {L{\"o}ffler}, {Lorca},
  {Manteiga}, {Marchal}, {Marrese}, {Moitinho}, {Mora}, {Muinonen}, {Osborne},
  {Pancino}, {Pauwels}, {Petit}, {Recio-Blanco}, {Richards}, {Riello},
  {Rimoldini}, {Robin}, {Roegiers}, {Rybizki}, {Sarro}, {Siopis}, {Smith},
  {Sozzetti}, {Ulla}, {Utrilla}, {van Leeuwen}, {van Reeven}, {Abbas}, {Abreu
  Aramburu}, {Accart}, {Aerts}, {Aguado}, {Ajaj}, {Altavilla}, {{\'A}lvarez},
  {{\'A}lvarez Cid-Fuentes}, {Alves}, {Anderson}, {Anglada Varela}, {Antoja},
  {Audard}, {Baines}, {Baker}, {Balaguer-N{\'u}{\~n}ez}, {Balbinot}, {Balog},
  {Barache}, {Barbato}, {Barros}, {Barstow}, {Bartolom{\'e}}, {Bassilana},
  {Bauchet}, {Baudesson-Stella}, {Becciani}, {Bellazzini}, {Bernet}, {Bertone},
  {Bianchi}, {Blanco-Cuaresma}, {Boch}, {Bombrun}, {Bossini}, {Bouquillon},
  {Bragaglia}, {Bramante}, {Breedt}, {Bressan}, {Brouillet}, {Bucciarelli},
  {Burlacu}, {Busonero}, {Butkevich}, {Buzzi}, {Caffau}, {Cancelliere},
  {C{\'a}novas}, {Cantat-Gaudin}, {Carballo}, {Carlucci}, {Carnerero},
  {Carrasco}, {Casamiquela}, {Castellani}, {Castro-Ginard}, {Castro Sampol},
  {Chaoul}, {Charlot}, {Chemin}, {Chiavassa}, {Cioni}, {Comoretto}, {Cooper},
  {Cornez}, {Cowell}, {Crifo}, {Crosta}, {Crowley}, {Dafonte}, {Dapergolas},
  {David}, {David}, {de Laverny}, {De Luise}, {De March}, {De Ridder}, {de
  Souza}, {de Teodoro}, {de Torres}, {del Peloso}, {del Pozo}, {Delbo},
  {Delgado}, {Delgado}, {Delisle}, {Di Matteo}, {Diakite}, {Diener},
  {Distefano}, {Dolding}, {Eappachen}, {Edvardsson}, {Enke}, {Esquej}, {Fabre},
  {Fabrizio}, {Faigler}, {Fedorets}, {Fernique}, {Fienga}, {Figueras},
  {Fouron}, {Fragkoudi}, {Fraile}, {Franke}, {Gai}, {Garabato},
  {Garcia-Gutierrez}, {Garc{\'\i}a-Torres}, {Garofalo}, {Gavras}, {Gerlach},
  {Geyer}, {Giacobbe}, {Gilmore}, {Girona}, {Giuffrida}, {Gomel}, {Gomez},
  {Gonzalez-Santamaria}, {Gonz{\'a}lez-Vidal}, {Granvik},
  {Guti{\'e}rrez-S{\'a}nchez}, {Guy}, {Hauser}, {Haywood}, {Helmi}, {Hidalgo},
  {Hilger}, {H{\l}adczuk}, {Hobbs}, {Holland}, {Huckle}, {Jasniewicz},
  {Jonker}, {Juaristi Campillo}, {Julbe}, {Karbevska}, {Kervella}, {Khanna},
  {Kochoska}, {Kontizas}, {Kordopatis}, {Korn}, {Kostrzewa-Rutkowska},
  {Kruszy{\'n}ska}, {Lambert}, {Lanza}, {Lasne}, {Le Campion}, {Le Fustec},
  {Lebreton}, {Lebzelter}, {Leccia}, {Leclerc}, {Lecoeur-Taibi}, {Liao},
  {Licata}, {Lindstr{\o}m}, {Lister}, {Livanou}, {Lobel}, {Madrero Pardo},
  {Managau}, {Mann}, {Marchant}, {Marconi}, {Marcos Santos}, {Marinoni},
  {Marocco}, {Marshall}, {Martin Polo}, {Mart{\'\i}n-Fleitas}, {Masip},
  {Massari}, {Mastrobuono-Battisti}, {Mazeh}, {McMillan}, {Messina},
  {Michalik}, {Millar}, {Mints}, {Molina}, {Molinaro}, {Moln{\'a}r},
  {Montegriffo}, {Mor}, {Morbidelli}, {Morel}, {Morris}, {Mulone}, {Munoz},
  {Muraveva}, {Murphy}, {Musella}, {Noval}, {Ord{\'e}novic}, {Orr{\`u}},
  {Osinde}, {Pagani}, {Pagano}, {Palaversa}, {Palicio}, {Panahi}, {Pawlak},
  {Pe{\~n}alosa Esteller}, {Penttil{\"a}}, {Piersimoni}, {Pineau}, {Plachy},
  {Plum}, {Poggio}, {Poretti}, {Poujoulet}, {Pr{\v{s}}a}, {Pulone}, {Racero},
  {Ragaini}, {Rainer}, {Raiteri}, {Rambaux}, {Ramos}, {Ramos-Lerate}, {Re
  Fiorentin}, {Regibo}, {Reyl{\'e}}, {Ripepi}, {Riva}, {Rixon}, {Robichon},
  {Robin}, {Roelens}, {Rohrbasser}, {Romero-G{\'o}mez}, {Rowell}, {Royer},
  {Rybicki}, {Sadowski}, {Sagrist{\`a} Sell{\'e}s}, {Sahlmann}, {Salgado},
  {Salguero}, {Samaras}, {Sanchez Gimenez}, {Sanna}, {Santove{\~n}a},
  {Sarasso}, {Schultheis}, {Sciacca}, {Segol}, {Segovia}, {S{\'e}gransan},
  {Semeux}, {Shahaf}, {Siddiqui}, {Siebert}, {Siltala}, {Slezak}, {Smart},
  {Solano}, {Solitro}, {Souami}, {Souchay}, {Spagna}, {Spoto}, {Steele},
  {Steidelm{\"u}ller}, {Stephenson}, {S{\"u}veges}, {Szabados}, {Szegedi-Elek},
  {Taris}, {Tauran}, {Taylor}, {Teixeira}, {Thuillot}, {Tonello}, {Torra},
  {Torra}, {Turon}, {Unger}, {Vaillant}, {van Dillen}, {Vanel}, {Vecchiato},
  {Viala}, {Vicente}, {Voutsinas}, {Weiler}, {Wevers}, {Wyrzykowski}, {Yoldas},
  {Yvard}, {Zhao}, {Zorec}, {Zucker}, {Zurbach}, \&
  {Zwitter}}]{2021A&A...649A...1G}
{Gaia Collaboration}, {Brown}, A.~G.~A., {Vallenari}, A., {et~al.} 2021, \aap,
  649, A1

\bibitem[{{Genzel} {et~al.}(2010){Genzel}, {Tacconi}, {Gracia-Carpio},
  {Sternberg}, {Cooper}, {Shapiro}, {Bolatto}, {Bouch{\'e}}, {Bournaud},
  {Burkert}, {Combes}, {Comerford}, {Cox}, {Davis}, {F{\"o}rster Schreiber},
  {Garcia-Burillo}, {Lutz}, {Naab}, {Neri}, {Omont}, {Shapley}, \&
  {Weiner}}]{2010MNRAS.407.2091G}
{Genzel}, R., {Tacconi}, L.~J., {Gracia-Carpio}, J., {et~al.} 2010, \mnras,
  407, 2091

\bibitem[{{Genzel} {et~al.}(2015){Genzel}, {Tacconi}, {Lutz}, {Saintonge},
  {Berta}, {Magnelli}, {Combes}, {Garc{\'\i}a-Burillo}, {Neri}, {Bolatto},
  {Contini}, {Lilly}, {Boissier}, {Boone}, {Bouch{\'e}}, {Bournaud}, {Burkert},
  {Carollo}, {Colina}, {Cooper}, {Cox}, {Feruglio}, {F{\"o}rster Schreiber},
  {Freundlich}, {Gracia-Carpio}, {Juneau}, {Kovac}, {Lippa}, {Naab}, {Salome},
  {Renzini}, {Sternberg}, {Walter}, {Weiner}, {Weiss}, \&
  {Wuyts}}]{2015ApJ...800...20G}
{Genzel}, R., {Tacconi}, L.~J., {Lutz}, D., {et~al.} 2015, \apj, 800, 20

\bibitem[{{G{\'o}mez-Guijarro} {et~al.}(2019){G{\'o}mez-Guijarro}, {Riechers},
  {Pavesi}, {Magdis}, {Leung}, {Valentino}, {Toft}, {Aravena}, {Chapman},
  {Clements}, {Dannerbauer}, {Oliver}, {P{\'e}rez-Fournon}, \&
  {Valtchanov}}]{2019ApJ...872..117G}
{G{\'o}mez-Guijarro}, C., {Riechers}, D.~A., {Pavesi}, R., {et~al.} 2019, \apj,
  872, 117

\bibitem[{{Gullberg} {et~al.}(2019){Gullberg}, {Smail}, {Swinbank},
  {Dudzevi{\v{c}}i{\={u}}t{\.{e}}}, {Stach}, {Thomson}, {Almaini}, {Chen},
  {Conselice}, {Cooke}, {Farrah}, {Ivison}, {Maltby}, {Micha{\l}owski},
  {Simpson}, {Scott}, {Wardlow}, \& {Weiss}}]{2019MNRAS.490.4956G}
{Gullberg}, B., {Smail}, I., {Swinbank}, A.~M., {et~al.} 2019, \mnras, 490,
  4956

\bibitem[{{Harikane} {et~al.}(2019){Harikane}, {Ouchi}, {Ono}, {Fujimoto},
  {Donevski}, {Shibuya}, {Faisst}, {Goto}, {Hatsukade}, {Kashikawa}, {Kohno},
  {Hashimoto}, {Higuchi}, {Inoue}, {Lin}, {Martin}, {Overzier}, {Smail},
  {Toshikawa}, {Umehata}, {Ao}, {Chapman}, {Clements}, {Im}, {Jing},
  {Kawaguchi}, {Lee}, {Lee}, {Lin}, {Matsuoka}, {Marinello}, {Nagao},
  {Onodera}, {Toft}, \& {Wang}}]{2019ApJ...883..142H}
{Harikane}, Y., {Ouchi}, M., {Ono}, Y., {et~al.} 2019, \apj, 883, 142

\bibitem[{{Hayashi} {et~al.}(2011){Hayashi}, {Kodama}, {Koyama}, {Tadaki}, \&
  {Tanaka}}]{2011MNRAS.415.2670H}
{Hayashi}, M., {Kodama}, T., {Koyama}, Y., {Tadaki}, K.-I., \& {Tanaka}, I.
  2011, \mnras, 415, 2670

\bibitem[{{Hayashi} {et~al.}(2014){Hayashi}, {Kodama}, {Koyama}, {Tadaki},
  {Tanaka}, {Shimakawa}, {Matsuda}, {Sobral}, {Best}, \&
  {Smail}}]{2014MNRAS.439.2571H}
{Hayashi}, M., {Kodama}, T., {Koyama}, Y., {et~al.} 2014, \mnras, 439, 2571

\bibitem[{{Hayashi} {et~al.}(2017){Hayashi}, {Kodama}, {Kohno}, {Yamaguchi},
  {Tadaki}, {Hatsukade}, {Koyama}, {Shimakawa}, {Tamura}, \&
  {Suzuki}}]{2017ApJ...841L..21H}
{Hayashi}, M., {Kodama}, T., {Kohno}, K., {et~al.} 2017, \apjl, 841, L21

\bibitem[{{Hayashi} {et~al.}(2018){Hayashi}, {Tadaki}, {Kodama}, {Kohno},
  {Yamaguchi}, {Hatsukade}, {Koyama}, {Shimakawa}, {Tamura}, \&
  {Suzuki}}]{2018ApJ...856..118H}
{Hayashi}, M., {Tadaki}, K.-i., {Kodama}, T., {et~al.} 2018, \apj, 856, 118

\bibitem[{{Hilton} {et~al.}(2009){Hilton}, {Stanford}, {Stott}, {Collins},
  {Hoyle}, {Davidson}, {Hosmer}, {Kay}, {Liddle}, {Lloyd-Davies}, {Mann},
  {Mehrtens}, {Miller}, {Nichol}, {Romer}, {Sabirli}, {Sahl{\'e}n}, {Viana},
  {West}, {Barbary}, {Dawson}, {Meyers}, {Perlmutter}, {Rubin}, \&
  {Suzuki}}]{2009ApJ...697..436H}
{Hilton}, M., {Stanford}, S.~A., {Stott}, J.~P., {et~al.} 2009, \apj, 697, 436

\bibitem[{{Hilton} {et~al.}(2010){Hilton}, {Lloyd-Davies}, {Stanford}, {Stott},
  {Collins}, {Romer}, {Hosmer}, {Hoyle}, {Kay}, {Liddle}, {Mehrtens}, {Miller},
  {Sahl{\'e}n}, \& {Viana}}]{2010ApJ...718..133H}
{Hilton}, M., {Lloyd-Davies}, E., {Stanford}, S.~A., {et~al.} 2010, \apj, 718,
  133

\bibitem[{{Hine} {et~al.}(2016){Hine}, {Geach}, {Alexander}, {Lehmer},
  {Chapman}, \& {Matsuda}}]{2016MNRAS.455.2363H}
{Hine}, N.~K., {Geach}, J.~E., {Alexander}, D.~M., {et~al.} 2016, \mnras, 455,
  2363

\bibitem[{{Hodge} \& {da Cunha}(2020)}]{2020RSOS....700556H}
{Hodge}, J.~A., \& {da Cunha}, E. 2020, Royal Society Open Science, 7, 200556

\bibitem[{{Hodge} {et~al.}(2015){Hodge}, {Riechers}, {Decarli}, {Walter},
  {Carilli}, {Daddi}, \& {Dannerbauer}}]{2015ApJ...798L..18H}
{Hodge}, J.~A., {Riechers}, D., {Decarli}, R., {et~al.} 2015, \apjl, 798, L18

\bibitem[{{Hodge} {et~al.}(2016){Hodge}, {Swinbank}, {Simpson}, {Smail},
  {Walter}, {Alexander}, {Bertoldi}, {Biggs}, {Brandt}, {Chapman}, {Chen},
  {Coppin}, {Cox}, {Dannerbauer}, {Edge}, {Greve}, {Ivison}, {Karim},
  {Knudsen}, {Menten}, {Rix}, {Schinnerer}, {Wardlow}, {Weiss}, \& {van der
  Werf}}]{2016ApJ...833..103H}
{Hodge}, J.~A., {Swinbank}, A.~M., {Simpson}, J.~M., {et~al.} 2016, \apj, 833,
  103

\bibitem[{{Hopkins} {et~al.}(2006){Hopkins}, {Hernquist}, {Cox}, {Di Matteo},
  {Robertson}, \& {Springel}}]{2006ApJS..163....1H}
{Hopkins}, P.~F., {Hernquist}, L., {Cox}, T.~J., {et~al.} 2006, \apjs, 163, 1

\bibitem[{{Jaff{\'e}} {et~al.}(2015){Jaff{\'e}}, {Smith}, {Candlish},
  {Poggianti}, {Sheen}, \& {Verheijen}}]{2015MNRAS.448.1715J}
{Jaff{\'e}}, Y.~L., {Smith}, R., {Candlish}, G.~N., {et~al.} 2015, \mnras, 448,
  1715

\bibitem[{{Jin} {et~al.}(2021){Jin}, {Dannerbauer}, {Emonts}, {Serra}, {Lagos},
  {Thomson}, {Bassini}, {Lehnert}, {Allison}, {Champagne}, {Inderm{\"u}hle},
  {Norris}, {Seymour}, {Shimakawa}, {Casey}, {De Breuck}, {Drouart}, {Hatch},
  {Kodama}, {Koyama}, {Macgregor}, {Miley}, {Overzier},
  {P{\'e}rez-Mart{\'\i}nez}, {Rodr{\'\i}guez-Espinosa}, {R{\"o}ttgering},
  {S{\'a}nchez Portal}, \& {Ziegler}}]{2021A&A...652A..11J}
{Jin}, S., {Dannerbauer}, H., {Emonts}, B., {et~al.} 2021, \aap, 652, A11

\bibitem[{{Kaasinen} {et~al.}(2020){Kaasinen}, {Walter}, {Novak}, {Neeleman},
  {Smail}, {Boogaard}, {Cunha}, {Weiss}, {Liu}, {Decarli}, {Popping},
  {Diaz-Santos}, {Cort{\'e}s}, {Aravena}, {Werf}, {Riechers}, {Inami}, {Hodge},
  {Rix}, \& {Cox}}]{2020ApJ...899...37K}
{Kaasinen}, M., {Walter}, F., {Novak}, M., {et~al.} 2020, \apj, 899, 37

\bibitem[{{Kauffmann} {et~al.}(2004){Kauffmann}, {White}, {Heckman},
  {M{\'e}nard}, {Brinchmann}, {Charlot}, {Tremonti}, \&
  {Brinkmann}}]{2004MNRAS.353..713K}
{Kauffmann}, G., {White}, S. D.~M., {Heckman}, T.~M., {et~al.} 2004, \mnras,
  353, 713

\bibitem[{{Kaviraj} {et~al.}(2013){Kaviraj}, {Cohen}, {Windhorst}, {Silk},
  {O'Connell}, {Dopita}, {Dekel}, {Hathi}, {Straughn}, \&
  {Rutkowski}}]{2013MNRAS.429L..40K}
{Kaviraj}, S., {Cohen}, S., {Windhorst}, R.~A., {et~al.} 2013, \mnras, 429, L40

\bibitem[{{Kennicutt}(1989)}]{1989ApJ...344..685K}
{Kennicutt}, Robert~C., J. 1989, \apj, 344, 685

\bibitem[{{Kennicutt}(1998{\natexlab{a}})}]{1998ARA&A..36..189K}
---. 1998{\natexlab{a}}, \araa, 36, 189

\bibitem[{{Kennicutt}(1998{\natexlab{b}})}]{1998ApJ...498..541K}
---. 1998{\natexlab{b}}, \apj, 498, 541

\bibitem[{{Kennicutt} \& {de los Reyes}(2021)}]{2021ApJ...908...61K}
{Kennicutt}, Robert~C., J., \& {de los Reyes}, M. A.~C. 2021, \apj, 908, 61

\bibitem[{{Kodama} {et~al.}(2004){Kodama}, {Yamada}, {Akiyama}, {Aoki}, {Doi},
  {Furusawa}, {Fuse}, {Imanishi}, {Ishida}, {Iye}, {Kajisawa}, {Karoji},
  {Kobayashi}, {Komiyama}, {Kosugi}, {Maeda}, {Miyazaki}, {Mizumoto},
  {Morokuma}, {Nakata}, {Noumaru}, {Ogasawara}, {Ouchi}, {Sasaki}, {Sekiguchi},
  {Shimasaku}, {Simpson}, {Takata}, {Tanaka}, {Ueda}, {Yasuda}, \&
  {Yoshida}}]{2004MNRAS.350.1005K}
{Kodama}, T., {Yamada}, T., {Akiyama}, M., {et~al.} 2004, \mnras, 350, 1005

\bibitem[{{Kriek} {et~al.}(2009){Kriek}, {van Dokkum}, {Labb{\'e}}, {Franx},
  {Illingworth}, {Marchesini}, \& {Quadri}}]{2009ApJ...700..221K}
{Kriek}, M., {van Dokkum}, P.~G., {Labb{\'e}}, I., {et~al.} 2009, \apj, 700,
  221

\bibitem[{{Larson} {et~al.}(2016){Larson}, {Sanders}, {Barnes}, {Ishida},
  {Evans}, {U}, {Mazzarella}, {Kim}, {Privon}, {Mirabel}, \&
  {Flewelling}}]{2016ApJ...825..128L}
{Larson}, K.~L., {Sanders}, D.~B., {Barnes}, J.~E., {et~al.} 2016, \apj, 825,
  128

\bibitem[{{Lee} {et~al.}(2017){Lee}, {Tanaka}, {Kawabe}, {Kohno}, {Kodama},
  {Kajisawa}, {Yun}, {Nakanishi}, {Iono}, {Tamura}, {Hatsukade}, {Umehata},
  {Saito}, {Izumi}, {Aretxaga}, {Tadaki}, {Zeballos}, {Ikarashi}, {Wilson},
  {Hughes}, \& {Ivison}}]{2017ApJ...842...55L}
{Lee}, M.~M., {Tanaka}, I., {Kawabe}, R., {et~al.} 2017, \apj, 842, 55

\bibitem[{{Leroy} {et~al.}(2008){Leroy}, {Walter}, {Brinks}, {Bigiel}, {de
  Blok}, {Madore}, \& {Thornley}}]{2008AJ....136.2782L}
{Leroy}, A.~K., {Walter}, F., {Brinks}, E., {et~al.} 2008, \aj, 136, 2782

\bibitem[{{Leroy} {et~al.}(2013){Leroy}, {Walter}, {Sandstrom}, {Schruba},
  {Munoz-Mateos}, {Bigiel}, {Bolatto}, {Brinks}, {de Blok}, {Meidt}, {Rix},
  {Rosolowsky}, {Schinnerer}, {Schuster}, \& {Usero}}]{2013AJ....146...19L}
{Leroy}, A.~K., {Walter}, F., {Sandstrom}, K., {et~al.} 2013, \aj, 146, 19

\bibitem[{{Lin} {et~al.}(2019){Lin}, {Pan}, {Ellison}, {Belfiore}, {Shi},
  {S{\'a}nchez}, {Hsieh}, {Rowlands}, {Ramya}, {Thorp}, {Li}, \&
  {Maiolino}}]{2019ApJ...884L..33L}
{Lin}, L., {Pan}, H.-A., {Ellison}, S.~L., {et~al.} 2019, \apjl, 884, L33

\bibitem[{{Lin} {et~al.}(2020){Lin}, {Ellison}, {Pan}, {Thorp}, {Su},
  {S{\'a}nchez}, {Belfiore}, {Bothwell}, {Bundy}, {Chen}, {Concas}, {Hsieh},
  {Hsieh}, {Li}, {Maiolino}, {Masters}, {Newman}, {Rowlands}, {Shi},
  {Smethurst}, {Stark}, {Xiao}, \& {Yu}}]{2020ApJ...903..145L}
{Lin}, L., {Ellison}, S.~L., {Pan}, H.-A., {et~al.} 2020, \apj, 903, 145

\bibitem[{{Lotz} {et~al.}(2013){Lotz}, {Papovich}, {Faber}, {Ferguson},
  {Grogin}, {Guo}, {Kocevski}, {Koekemoer}, {Lee}, {McIntosh}, {Momcheva},
  {Rudnick}, {Saintonge}, {Tran}, {van der Wel}, \&
  {Willmer}}]{2013ApJ...773..154L}
{Lotz}, J.~M., {Papovich}, C., {Faber}, S.~M., {et~al.} 2013, \apj, 773, 154

\bibitem[{{Ma} {et~al.}(2015){Ma}, {Smail}, {Swinbank}, {Simpson}, {Thomson},
  {Chen}, {Danielson}, {Hilton}, {Tadaki}, {Stott}, \&
  {Kodama}}]{2015ApJ...806..257M}
{Ma}, C.~J., {Smail}, I., {Swinbank}, A.~M., {et~al.} 2015, \apj, 806, 257

\bibitem[{{Maier} {et~al.}(2019){Maier}, {Hayashi}, {Ziegler}, \&
  {Kodama}}]{2019A&A...626A..14M}
{Maier}, C., {Hayashi}, M., {Ziegler}, B.~L., \& {Kodama}, T. 2019, \aap, 626,
  A14

\bibitem[{{Mart{\'\i}-Vidal} {et~al.}(2014){Mart{\'\i}-Vidal}, {Vlemmings},
  {Muller}, \& {Casey}}]{2014A&A...563A.136M}
{Mart{\'\i}-Vidal}, I., {Vlemmings}, W.~H.~T., {Muller}, S., \& {Casey}, S.
  2014, \aap, 563, A136

\bibitem[{{Matharu} {et~al.}(2019){Matharu}, {Muzzin}, {Brammer}, {van der
  Burg}, {Auger}, {Hewett}, {van der Wel}, {van Dokkum}, {Balogh}, {Chan},
  {Demarco}, {Marchesini}, {Nelson}, {Noble}, {Wilson}, \&
  {Yee}}]{2019MNRAS.484..595M}
{Matharu}, J., {Muzzin}, A., {Brammer}, G.~B., {et~al.} 2019, \mnras, 484, 595

\bibitem[{{Matharu} {et~al.}(2020){Matharu}, {Muzzin}, {Brammer}, {van der
  Burg}, {Auger}, {Hewett}, {Chan}, {Demarco}, {van Dokkum}, {Marchesini},
  {Nelson}, {Noble}, \& {Wilson}}]{2020MNRAS.493.6011M}
---. 2020, \mnras, 493, 6011

\bibitem[{{McKinney} {et~al.}(2022){McKinney}, {Ramakrishnan}, {Lee}, {Pope},
  {Alberts}, {Chiang}, \& {Popescu}}]{2022ApJ...928...88M}
{McKinney}, J., {Ramakrishnan}, V., {Lee}, K.-S., {et~al.} 2022, \apj, 928, 88

\bibitem[{{McMullin} {et~al.}(2007){McMullin}, {Waters}, {Schiebel}, {Young},
  \& {Golap}}]{2007ASPC..376..127M}
{McMullin}, J.~P., {Waters}, B., {Schiebel}, D., {Young}, W., \& {Golap}, K.
  2007, in Astronomical Society of the Pacific Conference Series, Vol. 376,
  Astronomical Data Analysis Software and Systems XVI, ed. R.~A. {Shaw},
  F.~{Hill}, \& D.~J. {Bell}, 127

\bibitem[{{Miller} {et~al.}(2018){Miller}, {Chapman}, {Aravena}, {Ashby},
  {Hayward}, {Vieira}, {Wei{\ss}}, {Babul}, {B{\'e}thermin}, {Bradford},
  {Brodwin}, {Carlstrom}, {Chen}, {Cunningham}, {De Breuck}, {Gonzalez},
  {Greve}, {Harnett}, {Hezaveh}, {Lacaille}, {Litke}, {Ma}, {Malkan},
  {Marrone}, {Morningstar}, {Murphy}, {Narayanan}, {Pass}, {Perry}, {Phadke},
  {Rennehan}, {Rotermund}, {Simpson}, {Spilker}, {Sreevani}, {Stark},
  {Strandet}, \& {Strom}}]{2018Natur.556..469M}
{Miller}, T.~B., {Chapman}, S.~C., {Aravena}, M., {et~al.} 2018, \nat, 556, 469

\bibitem[{{Muzzin} {et~al.}(2012){Muzzin}, {Wilson}, {Yee}, {Gilbank},
  {Hoekstra}, {Demarco}, {Balogh}, {van Dokkum}, {Franx}, {Ellingson}, {Hicks},
  {Nantais}, {Noble}, {Lacy}, {Lidman}, {Rettura}, {Surace}, \&
  {Webb}}]{2012ApJ...746..188M}
{Muzzin}, A., {Wilson}, G., {Yee}, H.~K.~C., {et~al.} 2012, \apj, 746, 188

\bibitem[{{Nantais} {et~al.}(2016){Nantais}, {van der Burg}, {Lidman},
  {Demarco}, {Noble}, {Wilson}, {Muzzin}, {Foltz}, {DeGroot}, \&
  {Cooper}}]{2016A&A...592A.161N}
{Nantais}, J.~B., {van der Burg}, R. F.~J., {Lidman}, C., {et~al.} 2016, \aap,
  592, A161

\bibitem[{{Nantais} {et~al.}(2017){Nantais}, {Muzzin}, {van der Burg},
  {Wilson}, {Lidman}, {Foltz}, {DeGroot}, {Noble}, {Cooper}, \&
  {Demarco}}]{2017MNRAS.465L.104N}
{Nantais}, J.~B., {Muzzin}, A., {van der Burg}, R. F.~J., {et~al.} 2017,
  \mnras, 465, L104

\bibitem[{{Noble} {et~al.}(2017){Noble}, {McDonald}, {Muzzin}, {Nantais},
  {Rudnick}, {van Kampen}, {Webb}, {Wilson}, {Yee}, {Boone}, {Cooper},
  {DeGroot}, {Delahaye}, {Demarco}, {Foltz}, {Hayden}, {Lidman},
  {Manilla-Robles}, \& {Perlmutter}}]{2017ApJ...842L..21N}
{Noble}, A.~G., {McDonald}, M., {Muzzin}, A., {et~al.} 2017, \apjl, 842, L21

\bibitem[{{Noble} {et~al.}(2019){Noble}, {Muzzin}, {McDonald}, {Rudnick},
  {Matharu}, {Cooper}, {Demarco}, {Lidman}, {Nantais}, {van Kampen}, {Webb},
  {Wilson}, \& {Yee}}]{2019ApJ...870...56N}
{Noble}, A.~G., {Muzzin}, A., {McDonald}, M., {et~al.} 2019, \apj, 870, 56

\bibitem[{{Onodera} {et~al.}(2010){Onodera}, {Kuno}, {Tosaki}, {Kohno},
  {Nakanishi}, {Sawada}, {Muraoka}, {Komugi}, {Miura}, {Kaneko}, {Hirota}, \&
  {Kawabe}}]{2010ApJ...722L.127O}
{Onodera}, S., {Kuno}, N., {Tosaki}, T., {et~al.} 2010, \apjl, 722, L127

\bibitem[{{Oteo} {et~al.}(2018){Oteo}, {Ivison}, {Dunne}, {Manilla-Robles},
  {Maddox}, {Lewis}, {de Zotti}, {Bremer}, {Clements}, {Cooray}, {Dannerbauer},
  {Eales}, {Greenslade}, {Omont}, {Perez{\textendash}Fourn{\'o}n}, {Riechers},
  {Scott}, {van der Werf}, {Weiss}, \& {Zhang}}]{2018ApJ...856...72O}
{Oteo}, I., {Ivison}, R.~J., {Dunne}, L., {et~al.} 2018, \apj, 856, 72

\bibitem[{{Overzier}(2016)}]{2016A&ARv..24...14O}
{Overzier}, R.~A. 2016, \aapr, 24, 14

\bibitem[{{Peng} {et~al.}(2010){Peng}, {Ho}, {Impey}, \&
  {Rix}}]{2010AJ....139.2097P}
{Peng}, C.~Y., {Ho}, L.~C., {Impey}, C.~D., \& {Rix}, H.-W. 2010, \aj, 139,
  2097

\bibitem[{{Perret} {et~al.}(2014){Perret}, {Renaud}, {Epinat}, {Amram},
  {Bournaud}, {Contini}, {Teyssier}, \& {Lambert}}]{2014A&A...562A...1P}
{Perret}, V., {Renaud}, F., {Epinat}, B., {et~al.} 2014, \aap, 562, A1

\bibitem[{{Popping} {et~al.}(2022){Popping}, {Pillepich}, {Calistro Rivera},
  {Schulz}, {Hernquist}, {Kaasinen}, {Marinacci}, {Nelson}, \&
  {Vogelsberger}}]{2022MNRAS.510.3321P}
{Popping}, G., {Pillepich}, A., {Calistro Rivera}, G., {et~al.} 2022, \mnras,
  510, 3321

\bibitem[{{Puglisi} {et~al.}(2021){Puglisi}, {Daddi}, {Valentino}, {Magdis},
  {Liu}, {Kokorev}, {Circosta}, {Elbaz}, {Bournaud}, {Gomez-Guijarro}, {Jin},
  {Madden}, {Sargent}, \& {Swinbank}}]{2021MNRAS.508.5217P}
{Puglisi}, A., {Daddi}, E., {Valentino}, F., {et~al.} 2021, \mnras, 508, 5217

\bibitem[{{Romer} {et~al.}(2001){Romer}, {Viana}, {Liddle}, \&
  {Mann}}]{2001ApJ...547..594R}
{Romer}, A.~K., {Viana}, P. T.~P., {Liddle}, A.~R., \& {Mann}, R.~G. 2001,
  \apj, 547, 594

\bibitem[{{Rudnick} {et~al.}(2017){Rudnick}, {Hodge}, {Walter}, {Momcheva},
  {Tran}, {Papovich}, {da Cunha}, {Decarli}, {Saintonge}, {Willmer}, {Lotz}, \&
  {Lentati}}]{2017ApJ...849...27R}
{Rudnick}, G., {Hodge}, J., {Walter}, F., {et~al.} 2017, \apj, 849, 27

\bibitem[{{Rybak} {et~al.}(2019){Rybak}, {Calistro Rivera}, {Hodge}, {Smail},
  {Walter}, {van der Werf}, {da Cunha}, {Chen}, {Dannerbauer}, {Ivison},
  {Karim}, {Simpson}, {Swinbank}, \& {Wardlow}}]{2019ApJ...876..112R}
{Rybak}, M., {Calistro Rivera}, G., {Hodge}, J.~A., {et~al.} 2019, \apj, 876,
  112

\bibitem[{{Saintonge} {et~al.}(2017){Saintonge}, {Catinella}, {Tacconi},
  {Kauffmann}, {Genzel}, {Cortese}, {Dav{\'e}}, {Fletcher},
  {Graci{\'a}-Carpio}, {Kramer}, {Heckman}, {Janowiecki}, {Lutz}, {Rosario},
  {Schiminovich}, {Schuster}, {Wang}, {Wuyts}, {Borthakur}, {Lamperti}, \&
  {Roberts-Borsani}}]{2017ApJS..233...22S}
{Saintonge}, A., {Catinella}, B., {Tacconi}, L.~J., {et~al.} 2017, \apjs, 233,
  22

\bibitem[{{S{\'a}nchez-Garc{\'\i}a} {et~al.}(2022){S{\'a}nchez-Garc{\'\i}a},
  {Pereira-Santaella}, {Garc{\'\i}a-Burillo}, {Colina}, {Alonso-Herrero},
  {Villar-Mart{\'\i}n}, {Saito}, {D{\'\i}az-Santos}, {Piqueras L{\'o}pez},
  {Arribas}, {Bellocchi}, {Cazzoli}, \& {Labiano}}]{2022A&A...659A.102S}
{S{\'a}nchez-Garc{\'\i}a}, M., {Pereira-Santaella}, M., {Garc{\'\i}a-Burillo},
  S., {et~al.} 2022, \aap, 659, A102

\bibitem[{{Schmidt}(1959)}]{1959ApJ...129..243S}
{Schmidt}, M. 1959, \apj, 129, 243

\bibitem[{{Scudder} {et~al.}(2015){Scudder}, {Ellison}, {Momjian}, {Rosenberg},
  {Torrey}, {Patton}, {Fertig}, \& {Mendel}}]{2015MNRAS.449.3719S}
{Scudder}, J.~M., {Ellison}, S.~L., {Momjian}, E., {et~al.} 2015, \mnras, 449,
  3719

\bibitem[{{Scudder} {et~al.}(2012){Scudder}, {Ellison}, {Torrey}, {Patton}, \&
  {Mendel}}]{2012MNRAS.426..549S}
{Scudder}, J.~M., {Ellison}, S.~L., {Torrey}, P., {Patton}, D.~R., \& {Mendel},
  J.~T. 2012, \mnras, 426, 549

\bibitem[{{Shimakawa} {et~al.}(2014){Shimakawa}, {Kodama}, {Tadaki}, {Tanaka},
  {Hayashi}, \& {Koyama}}]{2014MNRAS.441L...1S}
{Shimakawa}, R., {Kodama}, T., {Tadaki}, K.~I., {et~al.} 2014, \mnras, 441, L1

\bibitem[{{Silva} {et~al.}(2018){Silva}, {Marchesini}, {Silverman}, {Skelton},
  {Iono}, {Martis}, {Marsan}, {Tadaki}, {Brammer}, \&
  {kartaltepe}}]{2018ApJ...868...46S}
{Silva}, A., {Marchesini}, D., {Silverman}, J.~D., {et~al.} 2018, \apj, 868, 46

\bibitem[{{Simons} {et~al.}(2021){Simons}, {Papovich}, {Momcheva}, {Trump},
  {Brammer}, {Estrada-Carpenter}, {Backhaus}, {Cleri}, {Finkelstein},
  {Giavalisco}, {Ji}, {Jung}, {Matharu}, \& {Weiner}}]{2021ApJ...923..203S}
{Simons}, R.~C., {Papovich}, C., {Momcheva}, I., {et~al.} 2021, \apj, 923, 203

\bibitem[{{Simpson} {et~al.}(2015){Simpson}, {Smail}, {Swinbank}, {Almaini},
  {Blain}, {Bremer}, {Chapman}, {Chen}, {Conselice}, {Coppin}, {Danielson},
  {Dunlop}, {Edge}, {Farrah}, {Geach}, {Hartley}, {Ivison}, {Karim}, {Lani},
  {Ma}, {Meijerink}, {Micha{\l}owski}, {Mortlock}, {Scott}, {Simpson},
  {Spaans}, {Thomson}, {van Kampen}, \& {van der Werf}}]{2015ApJ...799...81S}
{Simpson}, J.~M., {Smail}, I., {Swinbank}, A.~M., {et~al.} 2015, \apj, 799, 81

\bibitem[{{Smail} {et~al.}(2021){Smail}, {Dudzevi{\v{c}}i{\={u}}t{\.{e}}},
  {Stach}, {Almaini}, {Birkin}, {Chapman}, {Chen}, {Geach}, {Gullberg},
  {Hodge}, {Ikarashi}, {Ivison}, {Scott}, {Simpson}, {Swinbank}, {Thomson},
  {Walter}, {Wardlow}, \& {van der Werf}}]{2021MNRAS.502.3426S}
{Smail}, I., {Dudzevi{\v{c}}i{\={u}}t{\.{e}}}, U., {Stach}, S.~M., {et~al.}
  2021, \mnras, 502, 3426

\bibitem[{{Speagle} {et~al.}(2014){Speagle}, {Steinhardt}, {Capak}, \&
  {Silverman}}]{2014ApJS..214...15S}
{Speagle}, J.~S., {Steinhardt}, C.~L., {Capak}, P.~L., \& {Silverman}, J.~D.
  2014, \apjs, 214, 15

\bibitem[{{Spilker} {et~al.}(2019){Spilker}, {Bezanson}, {Weiner}, {Whitaker},
  \& {Williams}}]{2019ApJ...883...81S}
{Spilker}, J.~S., {Bezanson}, R., {Weiner}, B.~J., {Whitaker}, K.~E., \&
  {Williams}, C.~C. 2019, \apj, 883, 81

\bibitem[{{Stach} {et~al.}(2017){Stach}, {Swinbank}, {Smail}, {Hilton},
  {Simpson}, \& {Cooke}}]{2017ApJ...849..154S}
{Stach}, S.~M., {Swinbank}, A.~M., {Smail}, I., {et~al.} 2017, \apj, 849, 154

\bibitem[{{Stanford} {et~al.}(2006){Stanford}, {Romer}, {Sabirli}, {Davidson},
  {Hilton}, {Viana}, {Collins}, {Kay}, {Liddle}, {Mann}, {Miller}, {Nichol},
  {West}, {Conselice}, {Spinrad}, {Stern}, \& {Bundy}}]{2006ApJ...646L..13S}
{Stanford}, S.~A., {Romer}, A.~K., {Sabirli}, K., {et~al.} 2006, \apjl, 646,
  L13

\bibitem[{{Suess} {et~al.}(2019){Suess}, {Kriek}, {Price}, \&
  {Barro}}]{2019ApJ...877..103S}
{Suess}, K.~A., {Kriek}, M., {Price}, S.~H., \& {Barro}, G. 2019, \apj, 877,
  103

\bibitem[{{Tacchella} {et~al.}(2015){Tacchella}, {Carollo}, {Renzini},
  {F{\"o}rster Schreiber}, {Lang}, {Wuyts}, {Cresci}, {Dekel}, {Genzel},
  {Lilly}, {Mancini}, {Newman}, {Onodera}, {Shapley}, {Tacconi}, {Woo}, \&
  {Zamorani}}]{2015Sci...348..314T}
{Tacchella}, S., {Carollo}, C.~M., {Renzini}, A., {et~al.} 2015, Science, 348,
  314

\bibitem[{{Tacconi} {et~al.}(2020){Tacconi}, {Genzel}, \&
  {Sternberg}}]{2020ARA&A..58..157T}
{Tacconi}, L.~J., {Genzel}, R., \& {Sternberg}, A. 2020, \araa, 58, 157

\bibitem[{{Tadaki} {et~al.}(2017){Tadaki}, {Kodama}, {Nelson}, {Belli},
  {F{\"o}rster Schreiber}, {Genzel}, {Hayashi}, {Herrera-Camus}, {Koyama},
  {Lang}, {Lutz}, {Shimakawa}, {Tacconi}, {{\"U}bler}, {Wisnioski}, {Wuyts},
  {Hatsukade}, {Lippa}, {Nakanishi}, {Ikarashi}, {Kohno}, {Suzuki}, {Tamura},
  \& {Tanaka}}]{2017ApJ...841L..25T}
{Tadaki}, K.-i., {Kodama}, T., {Nelson}, E.~J., {et~al.} 2017, \apjl, 841, L25

\bibitem[{{Tadaki} {et~al.}(2019){Tadaki}, {Kodama}, {Hayashi}, {Shimakawa},
  {Koyama}, {Lee}, {Tanaka}, {Hatsukade}, {Iono}, {Kohno}, {Matsuda}, {Suzuki},
  {Tamura}, {Toshikawa}, \& {Umehata}}]{2019PASJ...71...40T}
{Tadaki}, K.-i., {Kodama}, T., {Hayashi}, M., {et~al.} 2019, \pasj, 71, 40

\bibitem[{{Tadaki} {et~al.}(2020){Tadaki}, {Belli}, {Burkert}, {Dekel},
  {F{\"o}rster Schreiber}, {Genzel}, {Hayashi}, {Herrera-Camus}, {Kodama},
  {Kohno}, {Koyama}, {Lee}, {Lutz}, {Mowla}, {Nelson}, {Renzini}, {Suzuki},
  {Tacconi}, {{\"U}bler}, {Wisnioski}, \& {Wuyts}}]{2020ApJ...901...74T}
{Tadaki}, K.-i., {Belli}, S., {Burkert}, A., {et~al.} 2020, \apj, 901, 74

\bibitem[{{Toshikawa} {et~al.}(2014){Toshikawa}, {Kashikawa}, {Overzier},
  {Shibuya}, {Ishikawa}, {Ota}, {Shimasaku}, {Tanaka}, {Hayashi}, {Niino}, \&
  {Onoue}}]{2014ApJ...792...15T}
{Toshikawa}, J., {Kashikawa}, N., {Overzier}, R., {et~al.} 2014, \apj, 792, 15

\bibitem[{{van der Wel} {et~al.}(2014){van der Wel}, {Franx}, {van Dokkum},
  {Skelton}, {Momcheva}, {Whitaker}, {Brammer}, {Bell}, {Rix}, {Wuyts},
  {Ferguson}, {Holden}, {Barro}, {Koekemoer}, {Chang}, {McGrath},
  {H{\"a}ussler}, {Dekel}, {Behroozi}, {Fumagalli}, {Leja}, {Lundgren},
  {Maseda}, {Nelson}, {Wake}, {Patel}, {Labb{\'e}}, {Faber}, {Grogin}, \&
  {Kocevski}}]{2014ApJ...788...28V}
{van der Wel}, A., {Franx}, M., {van Dokkum}, P.~G., {et~al.} 2014, \apj, 788,
  28

\bibitem[{{Walter} {et~al.}(2016){Walter}, {Decarli}, {Aravena}, {Carilli},
  {Bouwens}, {da Cunha}, {Daddi}, {Ivison}, {Riechers}, {Smail}, {Swinbank},
  {Weiss}, {Anguita}, {Assef}, {Bacon}, {Bauer}, {Bell}, {Bertoldi}, {Chapman},
  {Colina}, {Cortes}, {Cox}, {Dickinson}, {Elbaz}, {G{\'o}nzalez-L{\'o}pez},
  {Ibar}, {Inami}, {Infante}, {Hodge}, {Karim}, {Le Fevre}, {Magnelli}, {Neri},
  {Oesch}, {Ota}, {Popping}, {Rix}, {Sargent}, {Sheth}, {van der Wel}, {van der
  Werf}, \& {Wagg}}]{2016ApJ...833...67W}
{Walter}, F., {Decarli}, R., {Aravena}, M., {et~al.} 2016, \apj, 833, 67

\bibitem[{{Wang} {et~al.}(2018){Wang}, {Elbaz}, {Daddi}, {Liu}, {Kodama},
  {Tanaka}, {Schreiber}, {Zanella}, {Valentino}, {Sargent}, {Kohno}, {Xiao},
  {Pannella}, {Ciesla}, {Gobat}, \& {Koyama}}]{2018ApJ...867L..29W}
{Wang}, T., {Elbaz}, D., {Daddi}, E., {et~al.} 2018, \apjl, 867, L29

\bibitem[{{Wang} {et~al.}(2019){Wang}, {Schreiber}, {Elbaz}, {Yoshimura},
  {Kohno}, {Shu}, {Yamaguchi}, {Pannella}, {Franco}, {Huang}, {Lim}, \&
  {Wang}}]{2019Natur.572..211W}
{Wang}, T., {Schreiber}, C., {Elbaz}, D., {et~al.} 2019, \nat, 572, 211

\bibitem[{{Watson} {et~al.}(2019){Watson}, {Tran}, {Tomczak}, {Alcorn},
  {Salazar}, {Gupta}, {Momcheva}, {Papovich}, {van Dokkum}, {Brammer}, {Lotz},
  \& {Willmer}}]{2019ApJ...874...63W}
{Watson}, C., {Tran}, K.-V., {Tomczak}, A., {et~al.} 2019, \apj, 874, 63

\bibitem[{{Webb} {et~al.}(2017){Webb}, {Lowenthal}, {Yun}, {Noble}, {Muzzin},
  {Wilson}, {Yee}, {Cybulski}, {Aretxaga}, \& {Hughes}}]{2017ApJ...844L..17W}
{Webb}, T. M.~A., {Lowenthal}, J., {Yun}, M., {et~al.} 2017, \apjl, 844, L17

\bibitem[{{Whitaker} {et~al.}(2017){Whitaker}, {Pope}, {Cybulski}, {Casey},
  {Popping}, \& {Yun}}]{2017ApJ...850..208W}
{Whitaker}, K.~E., {Pope}, A., {Cybulski}, R., {et~al.} 2017, \apj, 850, 208

\bibitem[{{Williams} {et~al.}(2022){Williams}, {Alberts}, {Spilker}, {Noble},
  {Stefanon}, {Willmer}, {Bezanson}, {Narayanan}, \&
  {Whitaker}}]{2022ApJ...929...35W}
{Williams}, C.~C., {Alberts}, S., {Spilker}, J.~S., {et~al.} 2022, \apj, 929,
  35

\bibitem[{{Wuyts} {et~al.}(2016){Wuyts}, {Wisnioski}, {Fossati}, {F{\"o}rster
  Schreiber}, {Genzel}, {Davies}, {Mendel}, {Naab}, {R{\"o}ttgers}, {Wilman},
  {Wuyts}, {Bandara}, {Beifiori}, {Belli}, {Bender}, {Brammer}, {Burkert},
  {Chan}, {Galametz}, {Kulkarni}, {Lang}, {Lutz}, {Momcheva}, {Nelson},
  {Rosario}, {Saglia}, {Seitz}, {Tacconi}, {Tadaki}, {{\"U}bler}, \& {van
  Dokkum}}]{2016ApJ...827...74W}
{Wuyts}, E., {Wisnioski}, E., {Fossati}, M., {et~al.} 2016, \apj, 827, 74

\bibitem[{{Wuyts} {et~al.}(2010){Wuyts}, {Cox}, {Hayward}, {Franx},
  {Hernquist}, {Hopkins}, {Jonsson}, \& {van Dokkum}}]{2010ApJ...722.1666W}
{Wuyts}, S., {Cox}, T.~J., {Hayward}, C.~C., {et~al.} 2010, \apj, 722, 1666

\bibitem[{{Wuyts} {et~al.}(2011){Wuyts}, {F{\"o}rster Schreiber}, {Lutz},
  {Nordon}, {Berta}, {Altieri}, {Andreani}, {Aussel}, {Bongiovanni}, {Cepa},
  {Cimatti}, {Daddi}, {Elbaz}, {Genzel}, {Koekemoer}, {Magnelli}, {Maiolino},
  {McGrath}, {P{\'e}rez Garc{\'\i}a}, {Poglitsch}, {Popesso}, {Pozzi},
  {Sanchez-Portal}, {Sturm}, {Tacconi}, \& {Valtchanov}}]{2011ApJ...738..106W}
{Wuyts}, S., {F{\"o}rster Schreiber}, N.~M., {Lutz}, D., {et~al.} 2011, \apj,
  738, 106

\bibitem[{{Yamaguchi} {et~al.}(2019){Yamaguchi}, {Kohno}, {Hatsukade}, {Wang},
  {Yoshimura}, {Ao}, {Caputi}, {Dunlop}, {Egami}, {Espada}, {Fujimoto},
  {Hayatsu}, {Ivison}, {Kodama}, {Kusakabe}, {Nagao}, {Ouchi}, {Rujopakarn},
  {Tadaki}, {Tamura}, {Ueda}, {Umehata}, {Wang}, \&
  {Yun}}]{2019ApJ...878...73Y}
{Yamaguchi}, Y., {Kohno}, K., {Hatsukade}, B., {et~al.} 2019, \apj, 878, 73

\bibitem[{{Zavala} {et~al.}(2019){Zavala}, {Casey}, {Scoville}, {Champagne},
  {Chiang}, {Dannerbauer}, {Drew}, {Fu}, {Spilker}, {Spitler}, {Tran},
  {Treister}, \& {Toft}}]{2019ApJ...887..183Z}
{Zavala}, J.~A., {Casey}, C.~M., {Scoville}, N., {et~al.} 2019, \apj, 887, 183

\end{thebibliography}

\end{document}